\documentclass[fleqn,usenatbib]{mnras}

\usepackage[T1]{fontenc}

\DeclareRobustCommand{\VAN}[3]{#2}
\let\VANthebibliography\thebibliography
\def\thebibliography{\DeclareRobustCommand{\VAN}[3]{##3}\VANthebibliography}

\usepackage{graphicx}	
\usepackage{amsmath}	
\usepackage{amssymb}	
\usepackage{siunitx}
\usepackage{xcolor}
\usepackage{amsmath}
\usepackage{orcidlink}
\usepackage{newtxtext,newtxmath}

\title{A Multi-wavelength Study of Swift~J0503.7-2819: a Chimeric magnetic CV} 
\author[{\it{Swift}}~J0503.7-2819]{Kala .G. Pradeep$^{1}$\thanks{E-mail: kalagp17@gmail.com}\orcidlink{0000-0003-1905-2962},
Kulinder Pal Singh$^{1}$\thanks{kpsinghx52@gmail.com(KPS)}\orcidlink{0000-0001-6952-3887},
G. C. Dewangan$^{2}$\orcidlink{0000-0003-1589-2075},
Elias Aydi$^{3}$\orcidlink{0000-0001-8525-3442},
P. E. Barrett$^{4}$\orcidlink{0000-0002-8456-1424},
\newauthor
D. A. H. Buckley$^{5,6,7,8}$\orcidlink{0000-0002-7004-9956},
V. Girish$^{9}$,
K. L. Page$^{10}$\orcidlink{0000-0001-5624-2613},
S. B. Potter$^{5,12}$\orcidlink{0000-0002-5956-2249}, and 
E. M. Schlegel$^{11}$\orcidlink{0000-0002-4162-8190}
\\
$^{1}$Department of Physical Sciences, Indian Institute of Science Education and Research Mohali, Knowledge City, Sector 81, SAS Nagar, Punjab 140306, India\\
$^{2}$ The Inter-University Centre for Astronomy and Astrophysics (IUCAA)-Pune, India\\
$^{3}$Center for Data Intensive and Time Domain Astronomy, Department of Physics and Astronomy, Michigan State University, East Lansing, MI 48824, USA\\
$^{4}$Department of Physics, The George Washington University, 725 21st St. NW, Washington, DC 20052\\
$^{5}$South African Astronomical Observatory, PO Box 9, Observatory 7935, Cape Town, South Africa.\\
$^{6}$Southern African Large Telescope Pty. Ltd., SAAO, PO Box 9, Observatory 7935, Cape Town, South Africa.\\
$^{7}$Department of Astronomy, University of Cape Town, Private Bag X3, Rondebosch 7701, South Africa\\
$^{8}$Department of Physics, University of the Free State, PO Box 339, Bloemfontein 9300, South Africa\\
$^{9}$ Indian Space Research Organisation HQ, Antariksh Bhavan, New BEL Road, Bengaluru 560094, Karnataka, India\\
$^{10}$School of Physics and Astronomy, The University of Leicester, University Road, Leicester LE1 7RH, UK\\
$^{11}$The University of Texas at San Antonio,  One UTSA Circle, San Antonio, TX  78249, USA\\
$^{12}$Department of Physics, University of Johannesburg, PO Box 524, Auckland Park 2006, South Africa\\
}

\date{Accepted XXX. Received YYY; in original form ZZZ}

\pubyear{2023}

\begin{document}
\label{firstpage}
\pagerange{\pageref{firstpage}--\pageref{lastpage}}
\maketitle

\begin{abstract}
We present multi-wavelength temporal and spectral characteristics of a magnetic cataclysmic variable (MCV) Swift~J0503.7-2819, using far ultraviolet (FUV) and X-ray data from {\it{AstroSat}}, supplemented with optical data from the Southern African Large Telescope and X-ray data from the \textit{XMM-Newton} and \textit{Swift} observatories. The X-ray modulations at 4897.6657 s and 3932.0355 s are interpreted as the orbital ($P_{\Omega}$) and spin ($P_{\omega}$) period, respectively, and are consistent with prior reports. With a spin-orbit period ratio of 0.8 and $P_{\Omega}$ falling below the period gap (2-3 hrs) of CVs, Swift~J0503.7-2819 would be the newest addition to the growing population of nearly synchronous MCVs, which we call EX Hya-like systems. Hard X-ray luminosity of $<$ $2.5\times10^{32}~erg~s^{-1}$, as measured with the \textit{Swift} Burst Alert Telescope, identifies it to be a low-luminosity intermediate polar, similar to other EX~Hya-like systems. The phenomenology of the light curves and the spectral characteristics rule out a purely disc-fed/stream-fed model and instead reveal the presence of complex accretion structures around the white dwarf. We propose a ring-like accretion flow, akin to EX~Hya, using period ratio, stability arguments, and observational features. An attempt is made to differentiate between the asynchronous polar/nearly-synchronous intermediate polar nature of Swift~J0503.7-2819. Further, we note that with the advent of sensitive surveys, a growing population of MCVs that exhibit characteristics of both polars and intermediate polars is beginning to be identified, likely forming a genealogical link between the two conventional classes of MCVs. 
\end{abstract}

\begin{keywords}
accretion, accretion discs, (stars:) novae, cataclysmic variables, stars: individual: (SWIFT J0503.7-2819), stars: magnetic fields
\end{keywords}


\section{Introduction}

Cataclysmic Variables (CVs) are semi-detached binary systems with separations of a few solar radii and orbital periods of less than a day \citep[See,][for a detailed account of CVs.]{1995cvs..book.....W}. The binary is comprised of a white dwarf (WD) primary and a late-type main-sequence secondary, usually a red dwarf (RD), with the former accreting matter from the latter. The mass transfer in these systems is modeled as Roche-lobe overflow of the secondary. The geometry of the accretion flow depends on the binary separation and magnetic field strength of the primary and can take various forms viz., discs, rings, curtains, streams, or steam-overflow \citep[See,][for review of various accretion flows in CVs.]{2001cvs..book.....H, 2001LNP...573..155W}. In systems containing a rapidly spinning WD, the transferred matter gets magnetically propelled away from the primary, preventing accretion \citep[e.g.][]{1997MNRAS.286..436W}. Magnetic cataclysmic variables (MCVs) are CVs containing primaries with very strong magnetic fields ranging from $1-235$ MG. In MCVs, the WD is sufficiently magnetic to prevent the formation of a complete accretion disc. The extent and geometry of the magnetosphere that surrounds the WD dictate the mode by which accretion occurs in these magnetic systems. MCVs are broadly classified into Polars and Intermediate Polars (IPs).
Polars are characterized by a highly magnetic primary WD ($10-235$ MG). The close spatial proximity of the component stars facilitates the interaction of their magnetic fields; this, in conjunction with tidal locking, results in the coupling of the spin period ($P_{\omega}$) and the orbital period ($P_{\Omega}$). The matter in-flowing from the secondary accretes onto the primary, directly funneled by the magnetic field lines of the WD, forming an accretion stream. 
IPs consist of WDs with field strength ranging between 1-10 MG. The greater orbital separation of the constituent stars and the drag resulting from the interaction of their weak magnetic fields, together prevent spin-orbit synchronism. The spin and orbital periods of these MCVs are markedly asynchronous, with most IPs seen to attain spin equilibrium near  P$_{\omega}$/P$_{\Omega}$ $\sim$ 0.1 \citep{1999MNRAS.310..203K, 2000NewAR..44...75W}. The majority of IPs are located above the period gap of CVs i.e., $P_{\Omega}$ > $\approx 2 - 3$ hrs \citep{2010MNRAS.401.2207S}.
Due to its relatively weaker (compared to polars) magnetic field, the radius of the WD magnetosphere is smaller than those in a polar. Hence, the matter transferred from the secondary star is able to form a partial accretion disc before its inner regions are truncated by the WD's magnetic field, giving rise to the accretion structure called an accretion curtain. 
Asynchronous polars (APs) are a subgroup of polars that showcase diminutive asynchronism. Four MCVs are known to fall into this group, namely, V1500 Cyg, V1432 Aql, CD Ind, and BY Cam, all of them possessing a $P_{\Omega}$ value differing only by 2 \% or less from their respective $P_{\omega}$. The asynchronicity in these systems is understood to have arisen out of novae explosions suffered in the recent past. The secular spin period variation observed in these systems suggests they rapidly re-synchronize, with re-synchronization timescales ($t_{sync}$) in the order of several hundreds of years expected theoretically \citep{1987Ap&SS.131..557A}. Latest estimates of $t_{sync}$ of V1500 Cyg, V1432 Aql, CD Ind, and BY Cam are 150 yrs, 110 yrs, 6400 yrs and 1107 yrs, respectively \citep[See, Table 1 of][for a rundown of $t_{sync}$ estimates of APs over the years.]{2017PASP..129d4204M}. 
Yet another class of asynchronous MCVs with $P_{\Omega}$ $<$3 hrs and a $P_{\omega}/P_{\Omega}$ $>$ $0.1$ were classified by \cite{2004RMxAC..20..138N} as EX Hya-like systems. The eponymous EX Hya and other members of this group appear below the synchronization line, i.e., $P_{\omega}= P_{\Omega}$ in the $P_{\omega}-P_{\Omega}$ plane and deviate from the conventional IP spin equilibria, i.e., $P_{\omega}$/$P_{\Omega}$ $\sim$ 0.1. 

The emissions in MCVs extend throughout the electromagnetic spectrum. The transferred matter from the secondary (predominantly consisting of ionized Hydrogen) that accelerates along the field lines towards the surface of the WD generates an adiabatic shock just above the poles \citep{1990SSRv...54..195C}. The hot plasma in the post-shock region cools via thermal bremsstrahlung, radiating in the hard X-ray, and by cyclotron radiation in the optical and infrared if the magnetic field of the WD is significantly strong. The hard X-ray radiation emitted close to the WD is absorbed and re-radiated in the soft X-ray and FUV, as blackbody radiation. Optical light can also be produced in the disc (if present) as re-processed radiation. MCVs are ideal laboratories for studying accretion flows in extreme astrophysical environments. A multi-wavelength exploration can help gauge the emission properties across the accretion geometry which is crucial for understanding these structures. 

\vspace{.1cm}
Swift~J0503.7-2819 is a hard X-ray-selected MCV first observed by the {\it{Neil Gehrels Swift Observatory}} Burst Alert Telescope (BAT) all-sky hard X-ray survey \citep{2013ApJS..207...19B}. According to the GAIA EDR3 \citep{2016A&A...595A...1G,2021A&A...649A...1G} parallax of  1.142±0.086 milliarc-sec, it is at a distance of 875 parsecs. Based on time-series photometry and radial velocity spectroscopy, the system was initially characterized as an IP by \cite{2015AJ....150..170H}. They reported an orbital period of 4896$\pm4$ s from radial spectroscopy and time-series photometry, and the coherent oscillation at 975.2$\pm0.2$ s in the time-series was interpreted as the spin period of the system. With an orbital period below the period gap and an unusually large $P_{\omega}$/$P_{\Omega}$ value (0.2) for a hard X-ray selected IP, which is generally $<$ 0.1 \citep{2010MNRAS.401.2207S}, Swift~J0503.7-2819 showed a departure from conventional systems from the outset. Follow up X-ray time-series study by \cite{2022ApJ...934..123H} did not find a commensurate signal corresponding to the photometric spin period of 975 s in the {\it{XMM-Newton}} X-ray data, instead, two values for the spin period were reported, associated with two possible scenarios. A spin period of 3888 s is associated with the `one-pole/two-pole' picture (Model 1) while a 4337 s spin period is associated with the `pole-switching' model (Model 2). In both scenarios, the spin period is very close to the orbital period, such that the system was re-classified as an AP or a stream-fed IP. The X-ray temporal analysis of the same data reported by \cite{Rawat:2022onw} supports the case for the system being a stream-fed IP with a spin period of 3932 s, similar to Model 1 of \cite{2022ApJ...934..123H}. The optical spectroscopy carried out by \cite{2019Ap&SS.364..153M} suggests the magnetic activity in Swift~J0503.7-2819 is reminiscent of IPs. While the X-ray studies of the system point to disc-less accretion, the optical studies show the characteristics of systems with a disc. The system is particularly interesting considering there are several discrepancies when classifying it into conventional classes of MCVs and becomes pertinent in the context of the long-standing debate concerning the nature of accretion flow in asynchronous systems \citep[See section 1.5.3 of][for a summary of disc-ed/disc-less accretion debate]{Garlick_1993}. As pointed out by \cite{2023ApJ...943L..24L} long X-ray observations covering the beat cycle are essential to uncovering the accretion geometry of highly asynchronous systems such as Swift~J0503.7-2819, as the accretion flows in these tend to be modulated at the beat frequency. The {\it{AstroSat}}-SXT observations utilized in this paper cover the beat period of the system, including the longer of the two beat periods proposed by \cite{2022ApJ...934..123H}. The paper is organized as follows. In Section \ref{sec:obs_data} we detail the observations and the data reductions involved. The analysis of the reduced data and their results are reported in Section \ref{sec:Analysis}. The discussion and summary are presented in Section \ref{sec:Discussion} and \ref{sec:Summary}, respectively.

\section{Observations and Data Reduction}\label{sec:obs_data}

\begin{table*}
    \centering
    \caption {Log of Observations of Swift~J0503.7-2819. FUV filter or F1 = F148W of central wavelength 148 nm was used for UVIT observation. Refer, https://uvit.iiap.res.in/Instrument/Filters, for more information on UVIT filters. $^a$The BAT data were obtained from the 70 Month catalogue, so do not directly correspond to the Observation IDs listed. $^b$ Mean count rate for XRT spectrum comprising both Swift observations. $^c$ We have utilized ATLAS data from 2192 observations each of 30 s duration, the data are filtered to remove flux density values $<$ 0 and $>$ 1000 erg $cm^{-2}$ $s^{-1}$ and error values $>$ 100, the observation IDs correspond to the first and last observation in the series. $^d$  We have used 3745 TESS observations, each of 600 s.}
	 
    \label{tab:table1}
    \resizebox{2\columnwidth}{!}{
        \begin{tabular}{lccccl} 
        \hline
	Instrument  & Observation ID & Start Time (UT) & Stop Time (UT) & Exposure (s) & Count Rate \\
                   &  & YYYY-MM-DD HH:MM:SS & YYYY-MM-DD HH:MM:SS &  &  \\
		\hline
  Swift XRT+BAT$^a$   &  00041156001 & 2010-06-02 11:31:07 & 2010-06-02 22:54:55 &  6233 & $0.074 \pm 0.003^b$ \\
               & 00041156002    & 2010-06-09 04:31:24 & 2010-06-09 07:49:56 &  1775 & $0.074 \pm 0.003^b$ \\
  ATLAS$^c$ & 02a57303o0336o    & 2015-10-08 12:11:14 &         ...         &       &                     \\
            & 03a60189o0847o    &         ...         & 2023-09-02 02:41:40 & 65760 &   $241 \pm 33$      \\
  XMM-Newton   & 0801780301     & 2018-03-07 10:28:49 & 2018-03-07 18:15:29 & 28000 & $0.400 \pm 0.002$   \\
  AstroSat FUV F148W & 9000003776 & 2020-07-27 11:30:46 & 2020-07-31 13:00:59 & 20899 & $1.45 \pm 0.028$  \\
  SALT-RSS     & 2018-2-LSP-001 & 2020-07-31 03:55:28 & 2020-07-31 04:26:42 &  1800 & \\
  AstroSat SXT-1(0.35-5 keV) & 9000003776 & 2020-07-27 11:27:25 & 2020-07-31 19:31:34 & 41356 & $0.073 \pm 0.0035$ \\
  TESS$^d$  & TIC 686160823  & 2020-11-20 17:24:17 & 2020-12-16 17:34:10 &  2247000 & \\
  AstroSat SXT-2 (0.35-5 keV) & 9000005282 & 2022-08-06 09:51:35 & 2022-08-10 00:09:16 & 39578 & $0.085 \pm 0.0035$ \\
        \hline
	\end{tabular}
    }

\end{table*}

\subsection{{\it AstroSat Observations}}
{\it{AstroSat}} observed the source (Obs\_IDs: A09\_138T04\_900000x; x=3776/5282, PI: V. Girish) on two occasions. First observed on 2020 July, simultaneously using two co-aligned telescopes vis-{\` a}-vis, the Ultra-Violet Imaging Telescope (UVIT) \citep{2020AJ....159..158T} and the Soft X-ray Telescope (SXT) \citep{2017JApA...38...29S}, and the second, in 2022 August using SXT alone. A log of the observations is given in Table~\ref{tab:table1}, where the effective exposures and the mean count rates are also listed. 
UVIT is a three-in-one imaging telescope capable of observations in the near-UV (NUV), far-UV (FUV), and visible (VIS) ranges. It is constituted by two separate Ritchey-Chretien telescopes, each having a primary mirror with an effective radius of $\sim$187.5 mm. One of the twin telescopes is dedicated solely to observations in the FUV (130-180 nm) whereas the other, by employing a beam splitter, makes observations in both the NUV (200-300 nm) and the VIS (320-550 nm). UVIT has a field of view of 0.5 degrees and a spatial resolution (FWHM) of $\sim$1.5 arc-second.  
UVIT is operated only during orbital nights when it is occluded from the sun. As a result, the observations are interrupted at frequent intervals causing gaps in the data accumulated.  
The NUV and FUV channels are operated in Photon Counting/Centroiding (PC) mode, and the resulting data are lists of photon detection events, i.e., centroid-positions. The VIS channel operates in integration mode with the output being sequences of images. The VIS channel data are used for various corrections like flat-fielding, pointing drifts, etc., which occur during the course of observation and are not calibrated for scientific use.
UVIT observed Swift~J0503.7-2819 in FUV using the broadband $CaF_2$ filter centered at 148 nm, while the NUV channel, defunct since 2018, could not be used.

SXT is a focusing telescope working in the energy range of 0.3 to 8 keV using Wolter-I geometry. The X-ray photons are collected by a cooled CCD of 600$\times$600 pixels. The telescope has a FOV of 41.3 arc-minutes. SXT is capable of providing a spectral resolution of about 140 eV at 6 keV. It has a pointing accuracy of 0.5 arc-min and a spatial resolution of about 2 arc-min. SXT observed Swift~J0503.7-2819 on July 2020 and August 2022 (henceforth referred to as SXT-1 and SXT-2 observations, respectively). 

The reduction of data obtained from the two \textit{AstroSat} payloads are as follows: 

\textbf{AstroSat-UVIT:} The raw, Level 0 `L0’ data from UVIT are processed by the Indian Space Science Data Center (ISSDC) into Level 1 `L1’ and Level 2 `L2’ (end-user data; images, event lists, etc.) products. For the FUV image data reduction we have followed the \textsc{ccdlab} pipeline as elucidated in \cite{2017PASP..129k5002P}. \textsc{ccdlab}, a mission-specific software developed for UVIT image data reduction and astrometry uses L1 data obtained from the \textit{AstroSat} data repository, AstroBrowse, as input. The L1 data fed into the pipeline are converted into orbit-wise FITS images which are then corrected for various defects like pointing drifts, field distortions, and other instrumental artifacts. \textsc{ccdlab} uses the drift-series data obtained from VIS channel to correct for translational drift, the rotational drift between orbits is corrected by user-assisted registration of centroid lists to a common frame. The co-aligned orbit-wise images can now be stacked to improve the signal-to-noise (S/N) ratio, generating the deep field of the target. The next step in image data processing is astrometry. The deep field image is used to determine the World Coordinate Solution (WCS) for the field of view via astroquery \citep{2010AJ....139.1782L}, a novel method that uses trigonometric parameterization in conjunction with catalogue matching to solve for astronomical coordinates. The reduced FUV image was visualized using \textsc{SAO ds9} \citep{2000ascl.soft03002S} and the source counts were extracted from a circular region with radius of 40 pixels (1 pixel=0.4 arc-sec) centered on the source. The background counts were extracted from a rectangular region of size 100 px $\times$ 90 px from a neighbouring region bereft of any sources. The timing analysis was carried out using the UVIT-evt files generated from the Level 1 data. Transfiguring the UVIT-L1 data into event files makes them compatible with {\textsc{heasoft}} \citep{2014ascl.soft08004N}. To do this we used \textsc{ccdlab} to generate drift-corrected and aligned centroid list for each orbit. The centroid lists provide the positions of the detected photons in the detector. We converted these centroid lists to event files compatible with the \textsc{heasarc} tool {\textsc{xselect}}\footnote{\url{https://heasarc.gsfc.nasa.gov/ftools/xselect/}}, and merged them to a single event list for the entire observation.

\textbf{AstroSat-SXT:} The science products are obtained from the data reduction of Level 2 (L2) event files obtained from the \textit{AstroSat} archive. The L2 files are comprised of separate event files containing data of individual orbit-wise observations. The raw L1 data telemetered from the satellite to the data acquisition center are run through \textsc{sxtpipeline} to make them science-ready L2 data. The pipeline involves various processes vis-{\` a}-vis  calibration, event generation, time tagging, omission of spurious pixels etc., and is carried out using several ancillary files and tools, such as the calibration data-files provided by the instrument team and \textsc{sxtfilter} tool used for generating the Level-2 MKF filter file. Good Time Intervals (GTI) constructed to account for periods of inactivity (due to target occultation, solar illumination etc.) resulting in observation gaps are used to filter the data. The data once processed using the \textsc{sxtpipeline} are transformed into Level 2, clean event files \citep{2017JApA...38...29S}. The foremost step in L2 data processing is combining these individual event files into the merged, cleaned L2 event file and is carried out using the \textsc{sxtevtmerger tool}\footnote{The tool and its readme file can be downloaded from \textit{www.tifr.res.in/astrosat\_sxt/dataanalysis.}}. The tool merges the individual event files, after correcting for exposure overlaps within the optimum GTI. It cleans and combines the SXT orbit-wise event files into a file format that is compatible with \textsc{heasoft}. The merged data are screened using \textsc{xselect} for generating 
full-frame/higher order products that can be read by various utilities in \textsc{heasoft} such as \textsc{xspec}, \textsc{xronos} etc., with which the subsequent spectral and temporal analyses are carried out.

Soft X-ray image in the energy band of 0.35-5.0 keV was extracted from the merged event file using {\textsc {xselect}}.  
Based on radial profile inspection, a circular region of 12 arc-min radius and an annulus extending from 14 arc-min to 18 arc-min centered on the source were selected as the source and background regions respectively, to generate the science products.

\vspace{-0.3cm}
\subsection{{\it XMM-Newton Observations}}

\textit{Swift}~J0503.7-2819 was observed by \textit{XMM-Newton} on 2018 March 7 (ObsID 0801780301) for 28 ks. The processed event files were downloaded from the NASA Goddard Space Flight Center High Energy Astrophysics Science Archive Center (HEASARC). Only the EPIC MOS 1 data were used for the time series analysis.

\vspace{-0.3cm}
\subsection{{\it Swift Observations}}
{\it Swift} observed the source twice, in 2010 June, collecting a total of 8~ks of data with the X-ray Telescope (XRT). There is no evidence for a change in count rate between the two observations. A combined spectrum was therefore extracted, using the online \textsc{XRT product generator}\footnote{\url{https://www.swift.ac.uk/user\_objects/}} \citep{2009MNRAS.397.1177E}. In addition to the XRT, during the first {\it Swift} observation, on June 2, the UV/Optical Telescope observed the source with the uvw2 filter, measuring a Vega magnitude of uvw2 = $15.94 \pm 0.02$, while on June 9, the u filter was in place, leading to a measurement of u = $16.35 \pm 0.03$.

\subsection{{\it SALT Observations}}

We observed Swift~J0503.7-2819 on the night of 2020 July 31 (towards the end of the {\it AstroSat} observations) using the Robert Stobie Spectrograph \cite[RSS;][]{2003SPIE.4841.1463B, 2003SPIE.4841.1634K}, in Long Slit (LS) mode, mounted on the  Southern African Large Telescope  \cite[SALT;][]{2006SPIE.6267E..0ZB, 2006MNRAS.372..151O} situated at the South African Astronomical Observatory, Sutherland Station, South Africa. We obtained a time-series, 60\,s exposure, 30 spectra using the PG900 grating with a 1.5 arcsec slit, resulting in a resolution $R \sim 1000$ over the spectral range $4050 - 7100$\,\AA\AA.

The SALT spectra were first reduced using the \textsc{PySALT} pipeline \citep{2010SPIE.7737E..25C}, which involves bias subtraction, cross-talk correction, scattered light removal, bad pixel masking, and flat-fielding. The wavelength calibration, background subtraction, and spectral extraction are done using the \textsc{iraf} (Image Reduction and Analysis Facility) software \citep{1986SPIE..627..733T}.

\section{Analysis and Results}\label{sec:Analysis}

\begin{table*}
\caption{Optical and X-ray frequencies of Swift~J0503.7-2819. The frequencies in columns 2-7 are independent estimates from the combined \textit{ATLAS-TESS}, 
\textit{XMM-Newton} MOS~1, the two \textit{AstroSat} SXT periodograms - individually and combined; and the combined \textit{AstroSat} SXT and \textit{XMM-Newton}. Columns 8 and 9 show the measured and inferred frequencies and periods using the ATLAS-TESS data in Section \ref{subsec:X-ray_LCs_power}. The error in the frequencies is <1 in the least significant digit \citep[see][Equation 2.11, for the calculation of the standard deviation]{Bretthorst1988Bayesian}.}

\label{tab:frequencies}
\sisetup{table-alignment-mode=format, table-number-alignment=center}
\begin{tabular}{lS[table-format=1.8]S[table-format=1.4]S[table-format=1.4]S[table-format=1.4]S[table-format=1.7]S[table-format=1.7]S[table-format=1.8]S[table-format=5.5]}
\hline \hline
 Frequency       & \multicolumn{7}{c}{Frequency (mHz)} & {Period (s)} \\
Identification    & {ATLAS-TESS} & {MOS-1} & {SXT-1} & {SXT-2} & {SXT-1+SXT-2} & {MOS+SXT} & {Inferred} & {Inferred} \\
\hline
$\omega - \Omega$ &              & 0.0525  & 0.0498  & 0.0499  &   0.0498162   & 0.0491446 & 0.05014231 & 19943.23756 \\
$\Omega$          &   0.20417889 & 0.2034  &         &         &   0.2027129   & 0.2028688 & 0.20417889 &  4897.66596 \\
$\omega$          &   0.25432120 & 0.2564  & 0.2544  & 0.2538  &   0.2541670   & 0.2539793 & 0.25432120 &  3932.03555 \\
$2\Omega$         &   0.40829563 & 0.4043  &         &         &               &           &            &             \\
$\omega + \Omega$ &              & 0.4561  &         & 0.4585  &   0.4586770   & 0.4587547 & 0.45850009 &  2181.02465 \\
QPO 1             &              & 1.0443  & 1.3236  & 0.5004  &   1.4035112   &           &            &             \\
QPO 2             &              &         & 7.8668  & 4.1293  &   4.1299249   &           &            &             \\
\hline
\end{tabular}
\end{table*}

\subsection{Timing Analysis}

\subsubsection{FUV light curve and power spectra}
\begin{figure}
    \includegraphics[width=\columnwidth]{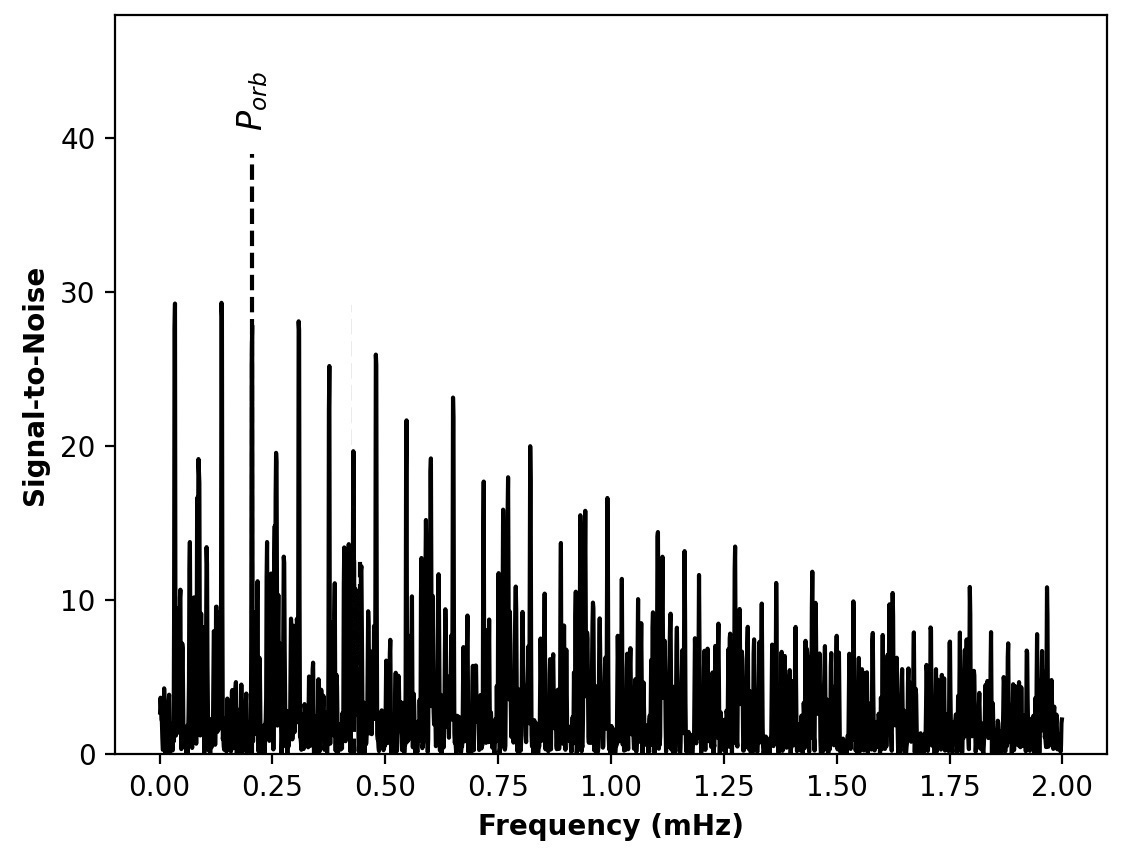}
    \caption{The initial Bretthorst periodogram of the \textit{AstroSat} UVIT FUV (F148W) data of Swift~J0503.7-2819. A potentially significant signal at 0.20499 mHz is identified with the orbital period of the system.}
    \label{fig:fuv-periodogram}
\end{figure}

The UVIT-FUV light curves extracted from the selected source and background region have average count rates of 1.450 counts s$^{-1}$ and 0.273 counts s$^{-1}$, respectively. After suitable rescaling, the background counts are subtracted from the source light curve yielding an average rate of 1.298 counts s$^{-1}$ for the background-subtracted curve. The initial Bretthorst periodogram analysis of the FUV light curve is shown in Figure \ref{fig:fuv-periodogram} \cite[for a description of the algorithm, see][ Appendix A.]{2021JApA...42...83S}. The periodogram shows a large number of aliased frequencies. To determine the correct frequencies, the latest
Asteroid Terrestrial-impact Last Alert System \cite[ATLAS;][]{2018PASP..130f4505T} and Transiting Exoplanet Survey Satellite \cite[TESS;][]{2015JATIS...1a4003R} data were combined into a single dataset and then analyzed to determine precise orbital and spin frequencies
 (see Table \ref{tab:frequencies}).
 The analysis proceeds as follows: First, the mean of the data is subtracted from the light curve to remove the constant offset or zero frequency; second, the periodogram is calculated and the most significant signal or the alias nearest to the known ATLAS+TESS frequency is subtracted from the light curve using the frequency and amplitude calculated from the periodogram. Step two is then repeated until there are no longer any significant frequencies. In Figure \ref{fig:fuv-periodogram} a potentially significant signal (at 0.20499 mHz) is marked. With an S/N $>$ 25 this peak corresponds to a period of 4878.1~s and is close to the previously reported orbital period of 4899 $\pm$ 2.4~s \citep{Rawat:2022onw}.  
Considering the FUV power spectrum is extremely noisy, the veracity of this signal is doubtful.

\begin{figure}
	\includegraphics[width=\columnwidth]{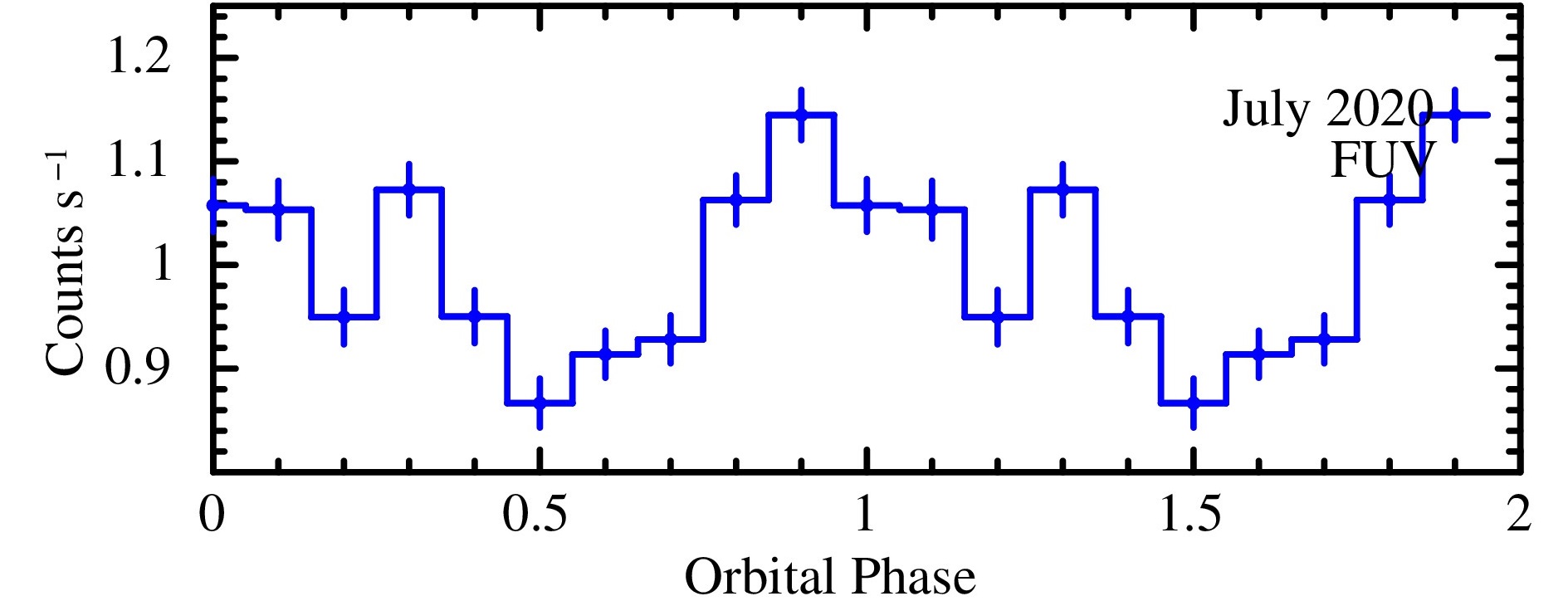}
       \caption{{\it{AstroSat}} FUV (F148W) orbital light curve ($P_{\Omega}$ = 4897.6657 s) of Swift~J0503.7-2819.
       Phase 0 is the epoch of blue-to-red crossing of the emission line radial velocity corresponding to BJD= 2456683.6274. 
       }
       
    \label{fig:FUV_SpinOrb_LCs}
\end{figure}

The FUV light curve is folded to the derived value of $P_{\Omega}$ 
and is shown in Figure \ref{fig:FUV_SpinOrb_LCs}. We have used the ephemeris from \cite{2022ApJ...934..123H}  with phase 0 corresponding to BJD 2456683.6274(9) for phase-folding the light curve. 
The orbital light curve displays a double-humped profile.  The curve shows
a less prominent first hump at orbital phase $\sim$0.3, peaks to an abrupt maximum at $\sim$0.9, and proceeds to more or less secularly decrease in count rate. The two peaks are separated by an interval of reduced emission rates which lasts for about 40 percent of the orbital cycle (from 0.4 to 0.8 phase). 
 
\subsubsection{ X-ray light curves and power spectra
}\label{subsec:X-ray_LCs_power}

\begin{figure*}
    \includegraphics[width=\textwidth]{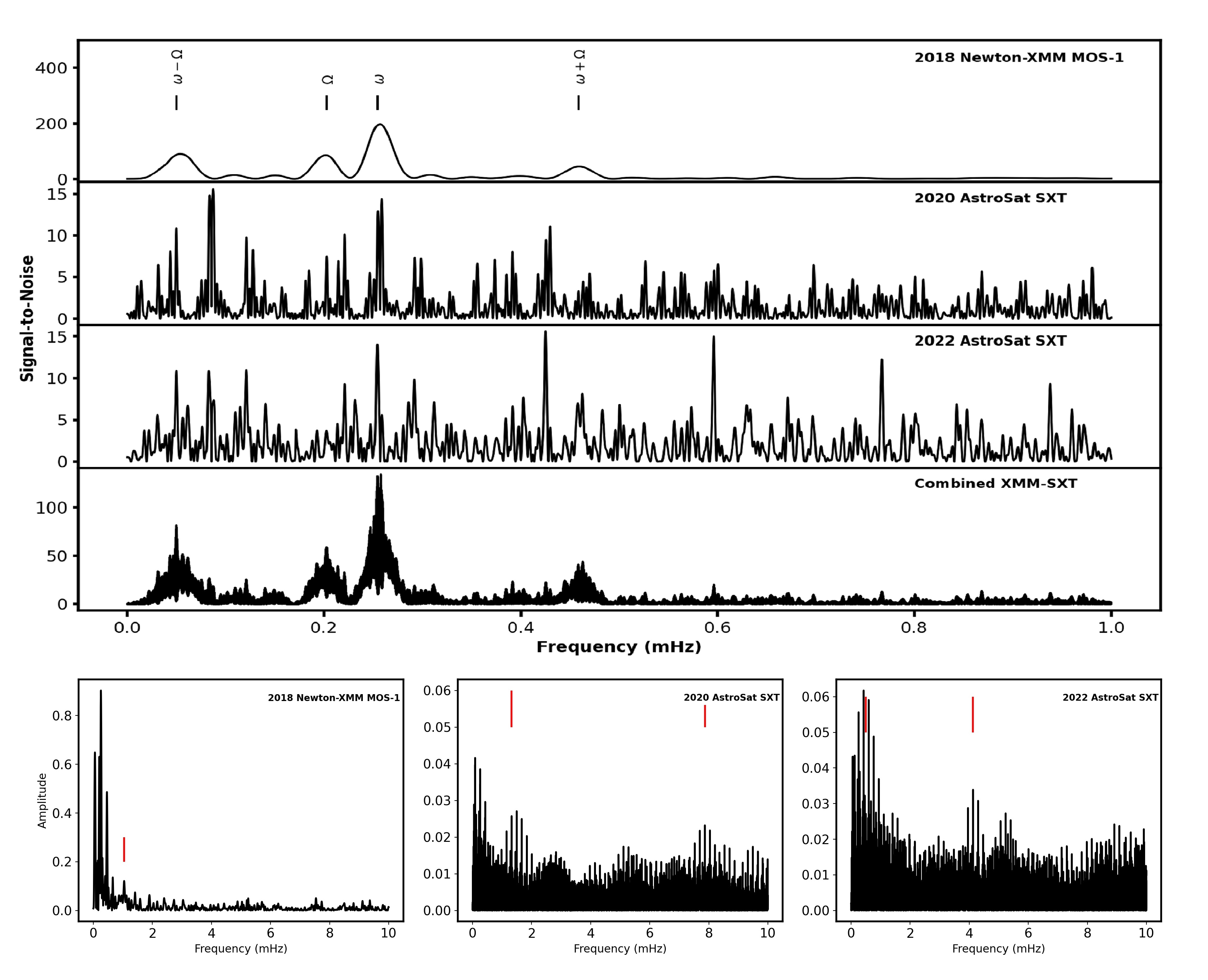}
    \caption{(Top Panel) The initial Bretthorst periodograms of the X-ray observations of Swift~J0503.7-2819. From top to bottom, the periodograms are of the 2018 \textit{XMM-Newton}, the 2020 and 2022 \textit{AstroSat} SXT, and the combined XMM-SXT data. The solid black line marks the frequencies of the orbital $\Omega$, spin $\omega$, and positive $\omega+\Omega$ and negative $\omega-\Omega$ beat frequencies. (Bottom Panel) The extended periodograms show the signals that could potentially be of a QPO and its harmonic, the central frequencies are denoted by solid red lines.}
    \label{fig:xray-periodograms}
\end{figure*}
\begin{figure*}
  \includegraphics[width=0.8\textwidth]{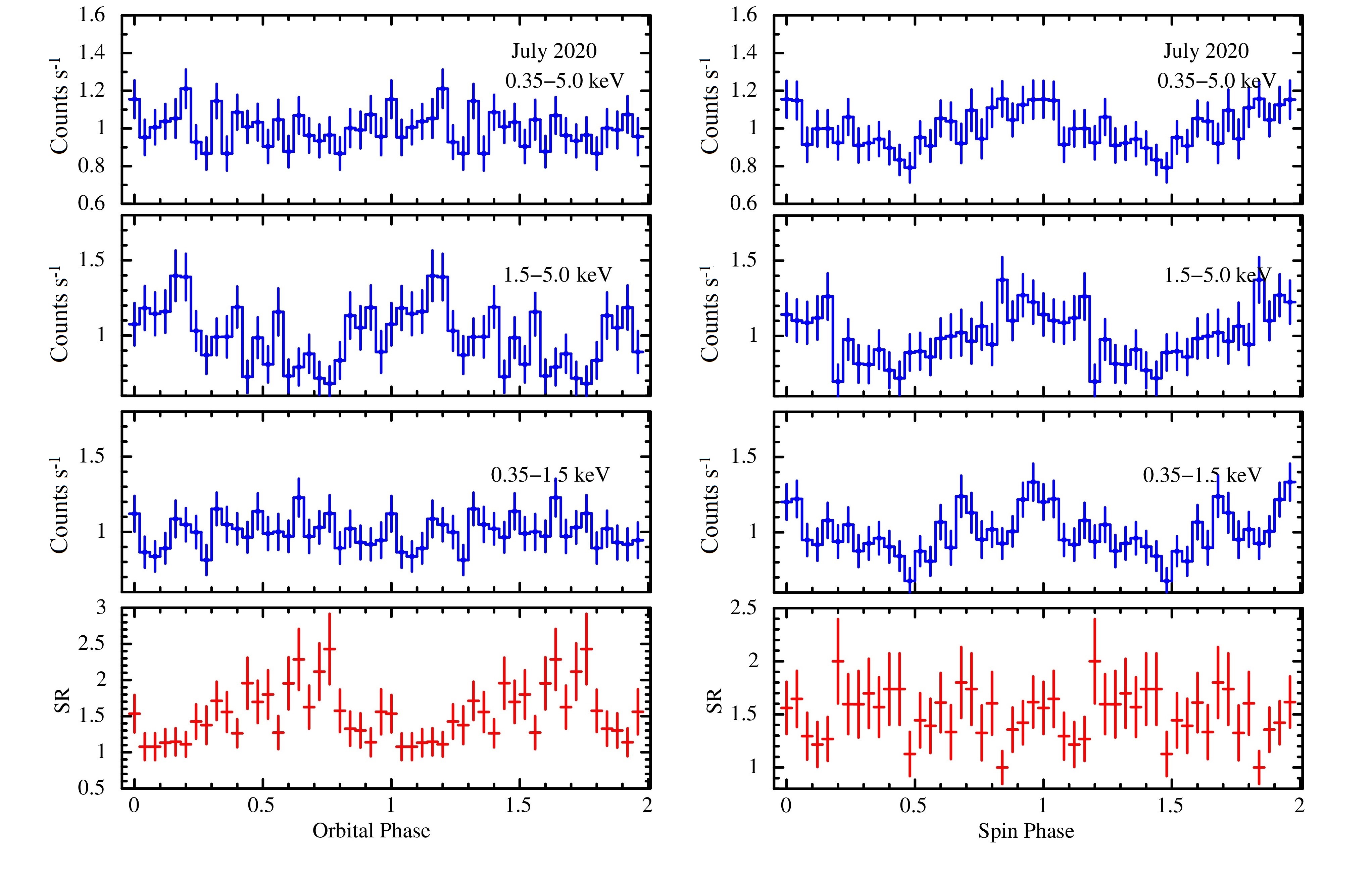}

  \includegraphics[width=0.8\textwidth]{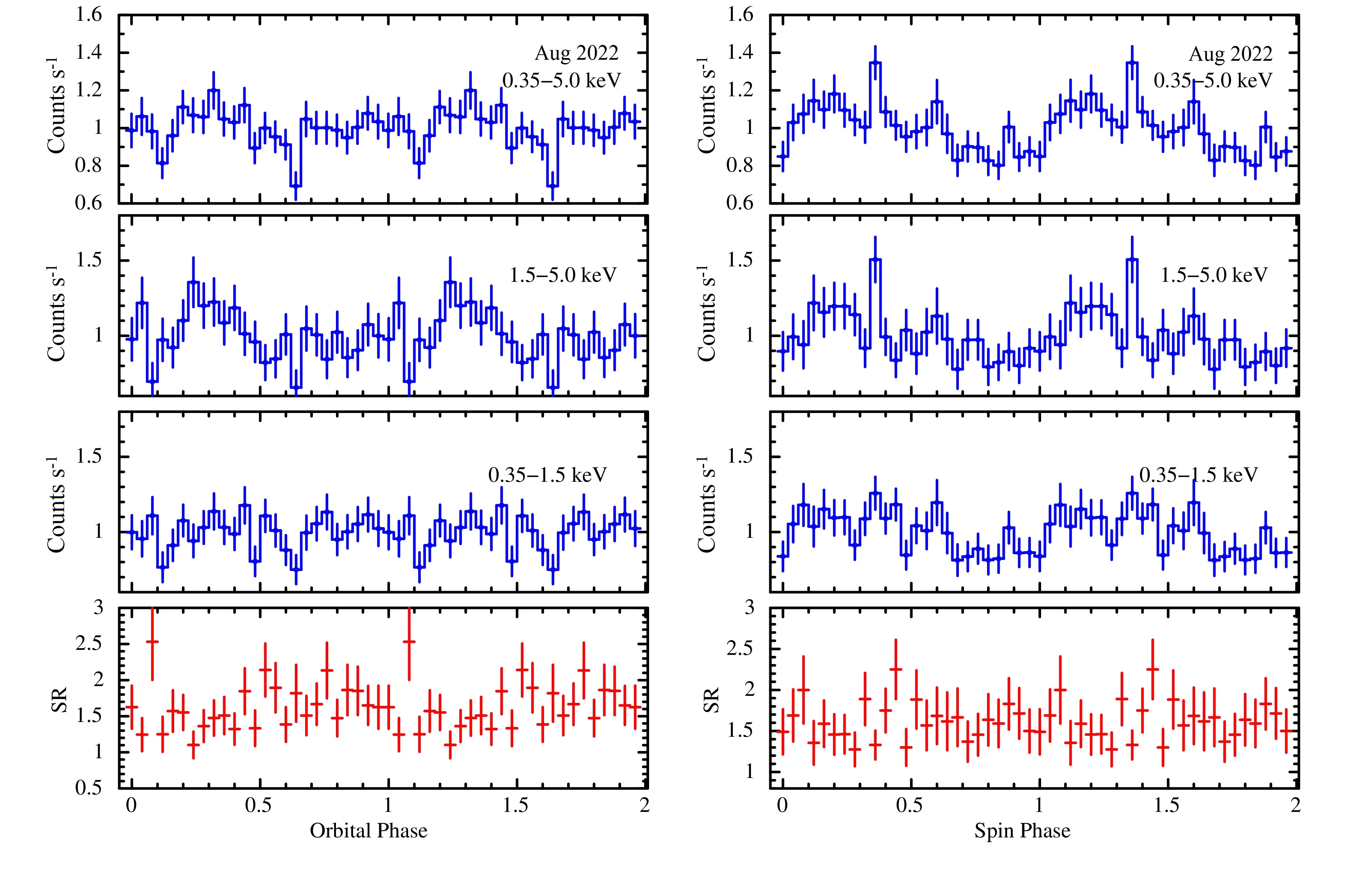}

     \caption{{\it{AstroSat}} SXT (0.3-5 keV) energy resolved orbit and spin phased light curves of Swift~J0503.7-2819 with softness ratio, SR = $\frac{0.35-1.5~keV}{1-1.5~keV}$. The phase-folded light curves from (Top) SXT-1 (July 2020) and (Bottom) SXT-2 (August 2022) are shown. Phase 0 of the orbital light curve is the epoch of blue-to-red crossing of the emission line radial velocity corresponding to BJD= 2456683.6274. The spin light curves have arbitrary phasing.}
   \label{fig:sxt_lc}
\end{figure*}

\begin{figure*}
    \includegraphics[width=6in]{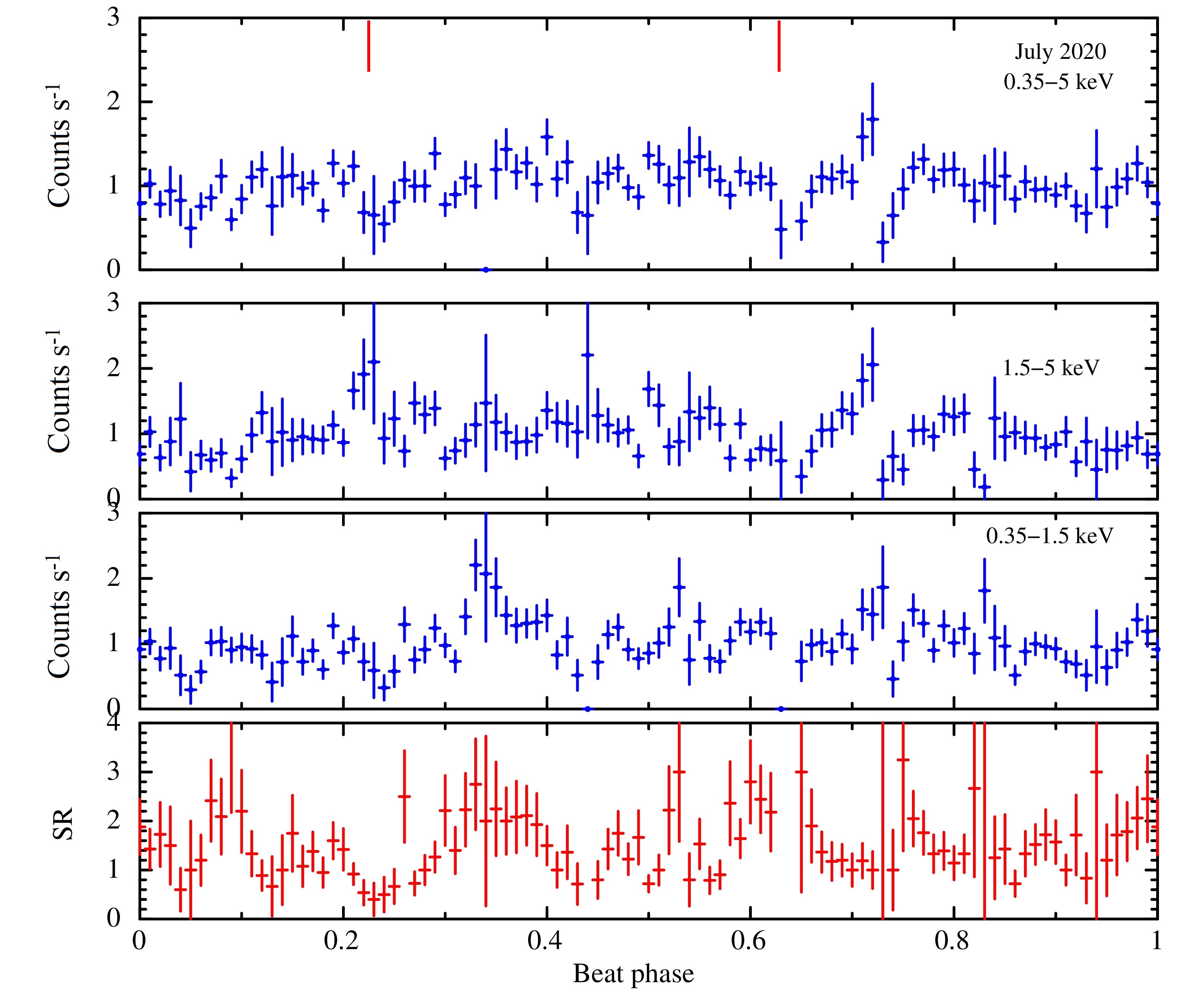}
    \caption{{(\it{AstroSat}} SXT (0.35-5) keV energy resolved light curve of Swift~J0503.7-2819 in 100s bins folded to beat period of 0.231 day. The phase range (0-1) covering a singular beat cycle is chosen so as to clearly discern the variations in emission characteristics/pole activity over the course of a beat period. The interval between the solid red lines roughly captures the duration of one-pole accretion. Phase 0 is arbitrary. 
    }
    \label{fig:swift_beat}
\end{figure*}

\begin{figure*}
      \includegraphics[width=6in]{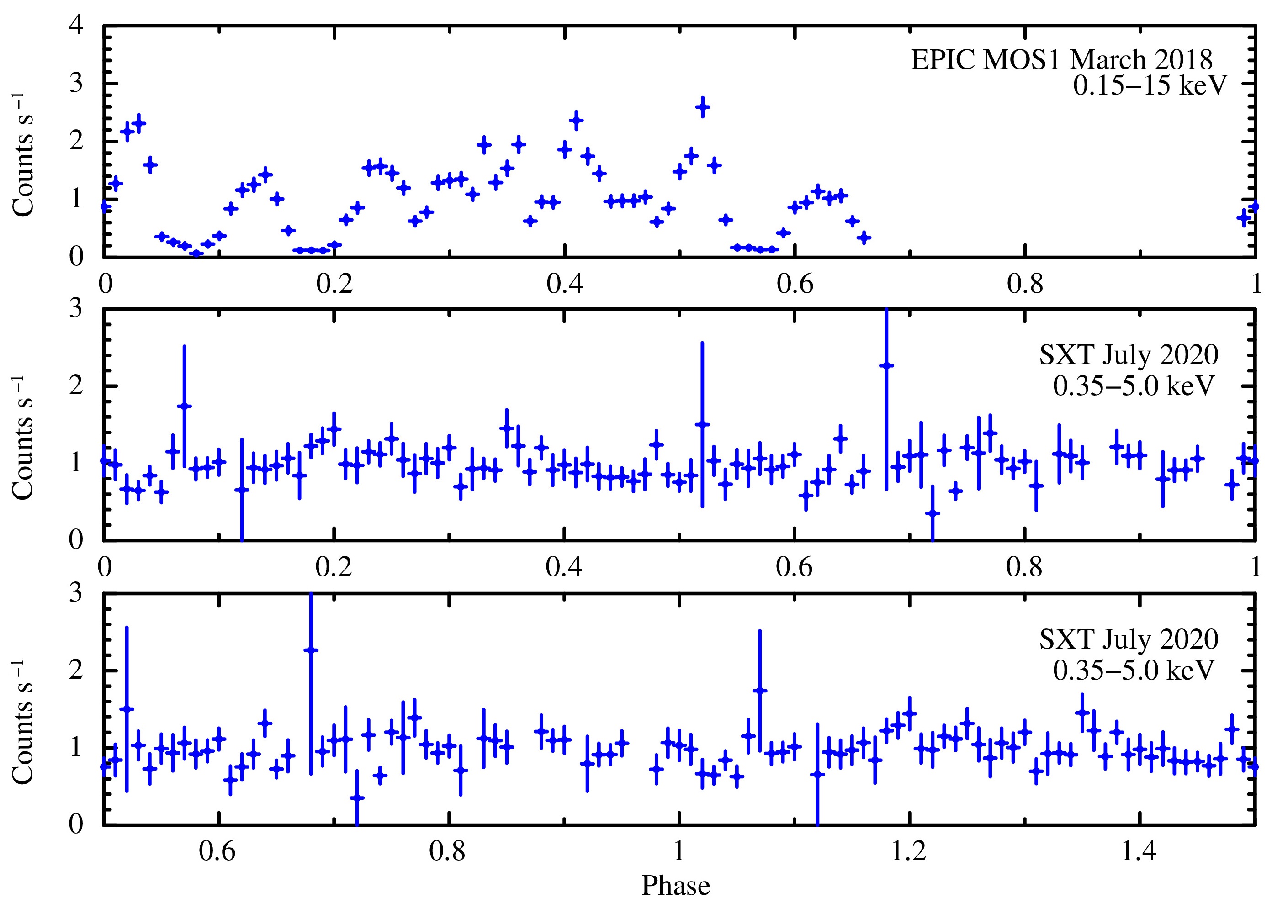}
    \caption{ X-ray light curves folded to `Model 2' beat period of 0.462 days proposed by \protect\cite{2022ApJ...934..123H}. (Top panel) {\it{XMM-Newton}} EPIC MOS 1 light curve, the {\it{AstroSat}} SXT-1 (0.35-5) keV light curve is plotted in the phase range 0-1 (Middle Panel) and 0.5-1.5 (Bottom Panel). The two segments of the SXT-1 light curve show similar morphology, i.e., the X-ray light curve has two identical half-beat cycles indicating similar pole activity throughout the proposed beat period hence a complete switching of the poles is discounted. The light curves have arbitrary phasing.}
    \label{fig:Model2_beat}
\end{figure*}
The soft X-ray time series data obtained from the two SXT observations, SXT-1 and 2, were extracted for the energy range of 0.35-5 keV from the selected source and background region. For SXT-1, the average count rate of the source (without background elimination) is estimated to be $0.73\times 10^{-1}$ counts s$^{-1}$ and the average count rate of the background is $0.43\times 10^{-1}$ counts s$^{-1}$. The re-scaled background counts are removed from the source, the resulting background-subtracted light curve has an average count rate of $0.26\times 10^{-1}$ counts s$^{-1}$. The same exercise was carried out for SXT-2 observation, the average count rate of the source, background, and background-substracted light curves are $0.85\times 10^{-1}$ counts s$^{-1}$, $0.44\times 10^{-1}$ counts s$^{-1}$ and $0.37\times 10^{-1}$ counts s$^{-1}$, respectively. Like the FUV periodogram, the X-ray periodograms also show a large number of aliased frequencies, because of the large gap between the two SXT observations and the gaps within each observation. To mitigate this situation, the MOS~1 data of the 2018 \textit{XMM-Newton} observation are also included in this analysis, because of its short, but continuous, observation. The \textit{XMM-Newton} analysis uses MOS~1 ($0.2-12$ keV) data in 20~s bins.  Processing Pipeline Subsystem (PPS) data were downloaded from the NASA Goddard Space Flight Center High Energy Science Archive Research Center (HEASARC) archive.  All X-ray data are barycentric corrected. Since the XMM data have a higher count rate than the SXT data (See Table \ref{tab:table1}), the XMM count rate is reduced by a factor of 3.3 so that the average count rate is similar to that of the SXT data. This is done to avoid injecting any spurious signals into the periodogram. Figure \ref{fig:xray-periodograms} shows the four initial periodograms. The top three panels are the individual periodograms of the MOS~1 and SXT observations, while the bottom panel is of the combined \textit{XMM-Newton} and \textit{AstroSat} X-ray data. The MOS~1 periodogram shows strong signals at the four most significant frequencies, which are believed to be the orbital $\Omega$, the spin $\omega$, and the positive $\omega + \Omega$ and negative $\omega - \Omega$ beat frequencies, whereas the SXT periodograms show only the spin and two beat frequencies. The identification of these frequencies is discussed in Section \ref{sec:Discussion}. As discussed in \cite{2022ApJ...934..123H}, the orbital frequency is more easily seen in hard X-rays than soft X-rays. However, it is seen, although weakly, in the combined SXT periodogram. These frequencies are listed in Table \ref{tab:frequencies} along with several additional significant frequenciss seen in the individual periodograms. These other frequencies are identified as the first harmonic of the spin frequency and quasi-periodic oscillations (QPOs) or superhumps. Note that the 755.5~s period (1.0443838 mHz) has a strong alias at 976~s, which is very close to the shortest period of 975~s seen in the optical MDM data \citep{2015AJ....150..170H}. The X-ray counterpart of this optical signal was elusive in the previous analyses of the \textit{XMM-Newton} data \cite[see, ][]{2022ApJ...934..123H, Rawat:2022onw}. The Lorenztian shape of the power spectral peak as opposed to a sharp one over a singular frequency is indicative of an underlying quasi-periodicity associated with the signal. The QPOs may also be resulting from the strong X-ray flaring clearly evident in the \textit{XMM-Newton} light curve \citep[See, Figure 1,][]{Rawat:2022onw}. It is, therefore, more likely that the signal at 975~s is that of QPO than a negative superhump. We also note that the latter is almost certain for short-period CVs with a low mass ratio \citep{1999dicb.conf...61P}. However, both interpretations presume structures in the orbital frame, whether these structures form a part of a disc or are long azimuth extensions of a stream, or connote an entirely different type of accretion structure can be ascertained in juxtaposition with other characteristics of the system. 

As shown in the Table \ref{tab:frequencies}, the X-ray periodograms do not provide a consistent and accurate set of frequencies for the orbital and spin frequencies by themselves. However, such frequencies can be estimated using the relatively accurate Asteroid Terrestrial-impact Last Alert System  \cite[ATLAS;][]{2018PASP..130f4505T} and TESS orbital and spin frequencies, respectively, as references. This is done by using the aliases of the TESS spin frequency and the positive X-ray beat frequency that give a frequency closest to the ATLAS orbital frequency. The inferred frequencies are listed in the last column of Table \ref{tab:frequencies}. 

The energy-resolved soft X-ray light curves from both SXT observations folded to the derived orbital and spin period are shown in Figure \ref{fig:sxt_lc}. The softness ratio (SR) is the ratio of counts in the energy range 0.35-1.5 keV and 1.5-5 keV. Both the orbital and spin-pulsed light curves show explicit energy dependence. The orbital modulation of the lower energy curve appears to be marginal with markedly lower counts compared to the higher energy curve. The indubitable orbital modulation of the SR curve shows there is significant photoelectric absorption by structures in the binary orbit. 

Figure \ref{fig:swift_beat} shows the energy-resolved light curve from the longer of the two SXT observations (SXT-1) along with the SR phased to the beat period. Due to the low S/N of the SXT light curve identifying modulation features, such as the individual spin pulses is a fraught process. However, it is clear that the system shows non-uniform emission characteristics during the course of the beat cycle. 
Between the phases $\sim$ 0.225-0.625, the variation of the light curve is quasi-sinusoidal and the behavior of the SR curve is distinctive from preceding and succeeding intervals with markedly dissimilar emission features in the soft and hard X-rays.  
A combination of self-occultation of the emitting region by the rotating WD and photo-electric absorption contributes to the drop in emission during this interval. Comparing the phenomenology of the SXT light curve with the shorter but high S/N light curve from {\it{XMM-Newton}} MOS1 (Figure \ref{fig:Model2_beat}), this interval of the SXT-1 light curve roughly captures the duration of one-pole accretion in the system, during which the `upper' pole is active while the `lower' pole is temporarily starved of accreting matter \cite[see][for more on the naming convention]{1999ASPC..157...33M}.
The intervals before and after One-pole accretion showcase sustained emission indicating an active, second, `lower' pole (henceforth referred to as `Two-pole' accretion).
This interval showcases near-equal count rates in the hard and soft X-rays throughout ($\sim$ 0.65 to the end of the cycle and till $\sim$ 0.25 from the beginning of the cycle). The depletion in emission in this interval is due to the self-occultation of WD as it spins. 

In Figure \ref{fig:Model2_beat} the {\it{XMM-Newton}} MOS 1 and {\it{AstroSat}} SXT-1 X-ray light curves are folded to the beat period associated with the `Pole-switching' model (Model 2) of \cite{2022ApJ...934..123H} which is twice as long as the beat period of One/Two-pole model i.e., 0.462 day. According to \cite{2022ApJ...934..123H}, in case of a complete pole-switching, the active pole changes over half-cycles i.e., the pole which is exclusively fed in the first half of the beat cycle will be starved the next, while the other pole becomes active and vice-versa. The beat-folded SXT light curve segments covering the intervals 0-1 and 0.5-1.5 show similar morphology. That is, the X-ray light curve folded to the beat period of 0.462 days has two identical half-beat cycles indicating similar pole activity throughout the proposed beat period thus suggesting that the pole which is continuously fed remains the same and accretion does not completely switch to the lower pole.

\subsection{Spectral Analysis}
\subsubsection{\textit{AstroSat SXT Spectra}}\label{subsec:SXT_Spectra}
\begin{figure*}
\includegraphics[width=0.6\textwidth]{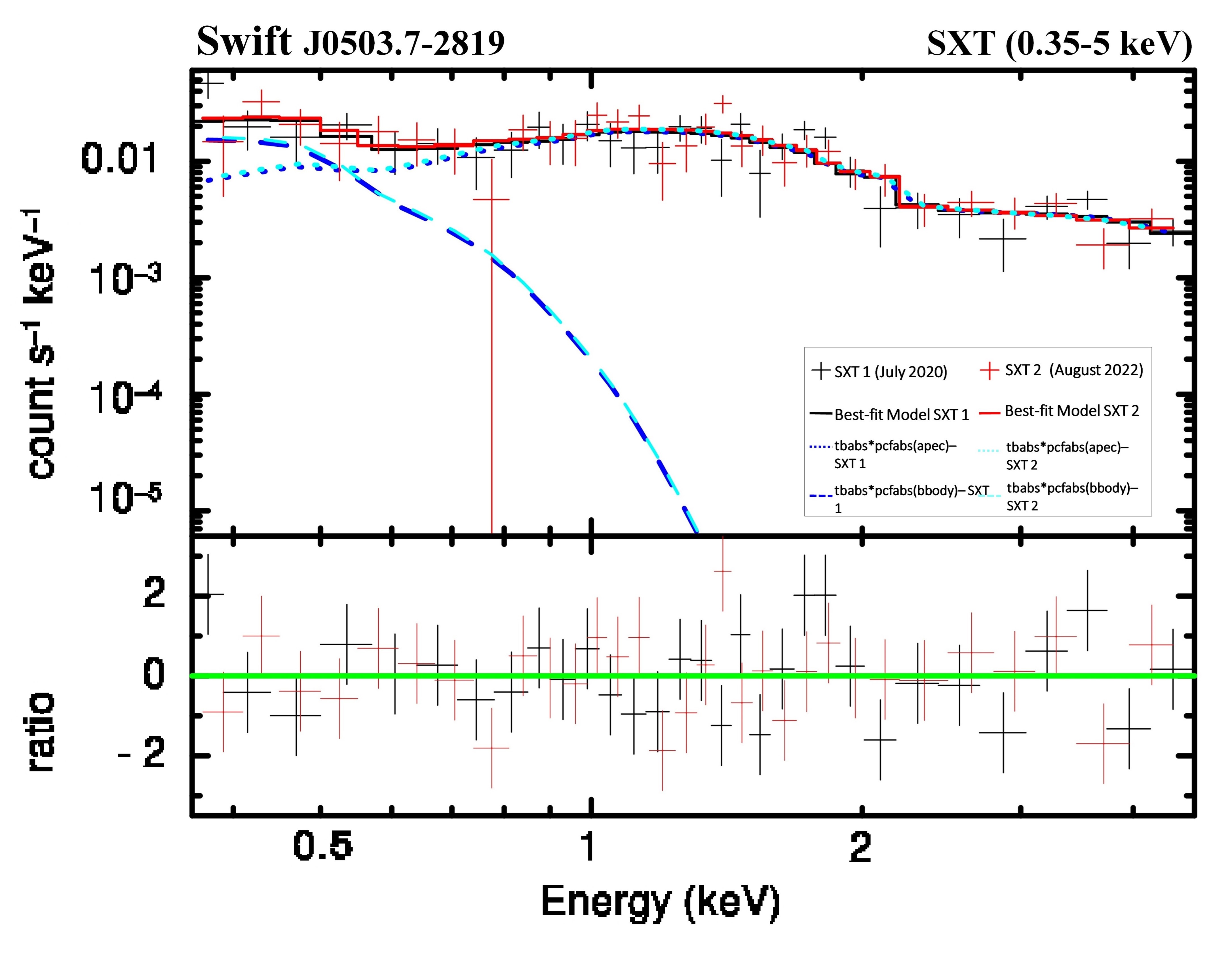}
    \caption{Soft X-ray joint spectra in 0.35-5.0 keV energy band of the SXT observations on July 2020 (SXT-1) and August 2022 (SXT-2). The best-fit spectral models for SXT-1 and SXT-2 are shown as black and red solid lines, respectively. The dashed lines show the contribution of the blackbody model and dotted lines show the contributions of the plasma emission model with SXT-1 represented in blue and SXT-2 in cyan. The residuals from the best-fit model are shown in the lower panel.}
   \label{fig:Swift_Spectra}
   \end{figure*}
The background-subtracted SXT soft X-ray spectra of Swift~J0503.7-2819 is shown in Figure \ref{fig:Swift_Spectra}. The spectra from SXT-1 and SXT-2 observations were extracted for the energy range of 0.35-5.0 keV and were fitted jointly using {\textsc{xspec}} version-12.12.0.  
The extracted data were analyzed using the latest SXT Response files from  \textit{www.tifr.res.in} vis-{\` a}-vis:\\
RMF File : $sxt\_pc\_mat\_g0to12.rmf$\\
ARF File : $sxt\_pc\_excl00\_v04\_20190608.arf$\\
The background file used for background elimination and scaling is sourced from the respective SXT observation so as to improve statistics. 

To understand the spectral properties of the system, different models (multiplicative and additive) and their combinations were adopted. A multiplicative model modifies the incident flux while an additive model embodies the emission source. The choice of the models was physically motivated by the characteristic emissions common in MCVs. The elemental abundance was set to that of solar using `{\it{aspl}}' \citep{2009ARA&A..47..481A}. To account for the absorption by the interstellar medium (ISM) the multiplicative model `{\it{tbabs}}' is used. `{\it{Tbabs}}' has a single associated parameter, the hydrogen column density `$N_{H}$' the value of which along the direction of Swift~J0503.7-2819 is $1.08 \times 10^{20} cm^{-2}$ \citep{2016A&A...594A.116H}, subsequent fits were carried out with this value of $N_{H}$, unchanged. The continuum emission is modeled using the additive models `\textit{bbody}' and `\textit{apec}'. The model `\textit{bbody}' accounts for the blackbody emission emanating from the white dwarf, while the model `{\it{apec}}'
accounts for the contribution of the collisionally ionized thermal plasma in the post-shock region. The spectral data are fit in a sequential fashion, identifying optimal values for each parameter of the models chosen. First, the spectra were fitted using two simple, single temperature models, sp., \textit{A = tbabs $\times$ bbody} and \textit{B = tbabs $\times$ apec}, both of which resulted in poor fits with unacceptably high values of the reduced $\chi^{2}$ ($\chi^{2}_{r}$). Multi-temperature fits were tried with the models {\it{C = tbabs $\times$ (apec+apec)}} and a combination of `{\it{bbody}}' and `{\it{apec}}', sp., {\it{D = tbabs $\times$ (bbody+apec)}}, both fits substantially brought down the $\chi^{2}_{r}$ to acceptable values (<2). Using an F-test, model `D' was identified to be more statistically significant than model `C', implying that the softest part of the spectrum is better modeled by a `\textit{bbody}' component than a soft `\textit{apec}' component. To refine the parametric values of `D' the blackbody temperature was kept frozen at the optimal value of 64.3 eV while the values of the temperature and normalization of the `\textit{apec}' model were allowed to vary. In most cases the `\textit{apec}' temperature was pegged at the upper limit of 64 keV, even the best of the `D' fit with a $\chi^{2}_{r}$ value of 1.18 could only bring down the temperature to 54 keV. Therefore, a new multiplicative model `{\it{pcfabs}}' that accounts for a partially covering absorber was introduced into the fit, with {\it{E = tbabs $\times$ pcfabs (bbody+apec)}}. The incorporation of `{\it{pcfabs}}' is physically meaningful as the X-ray emission in MCVs suffers local absorption by the accreting material. The `E' fit yielded an improved $\chi^{2}_{r}$ value of 1.05 and was able to constrain the {\it{apec}} temperature to 30 keV. The spectral parameters used for the spectral fits are listed in Table \ref{tab:Spectral_Fit_Table}. 

\subsubsection{{\it{Swift XRT+BAT Spectra}}}\label{subsec:XRT_Spectra}
   \begin{figure*}
\includegraphics[width=0.6\textwidth]{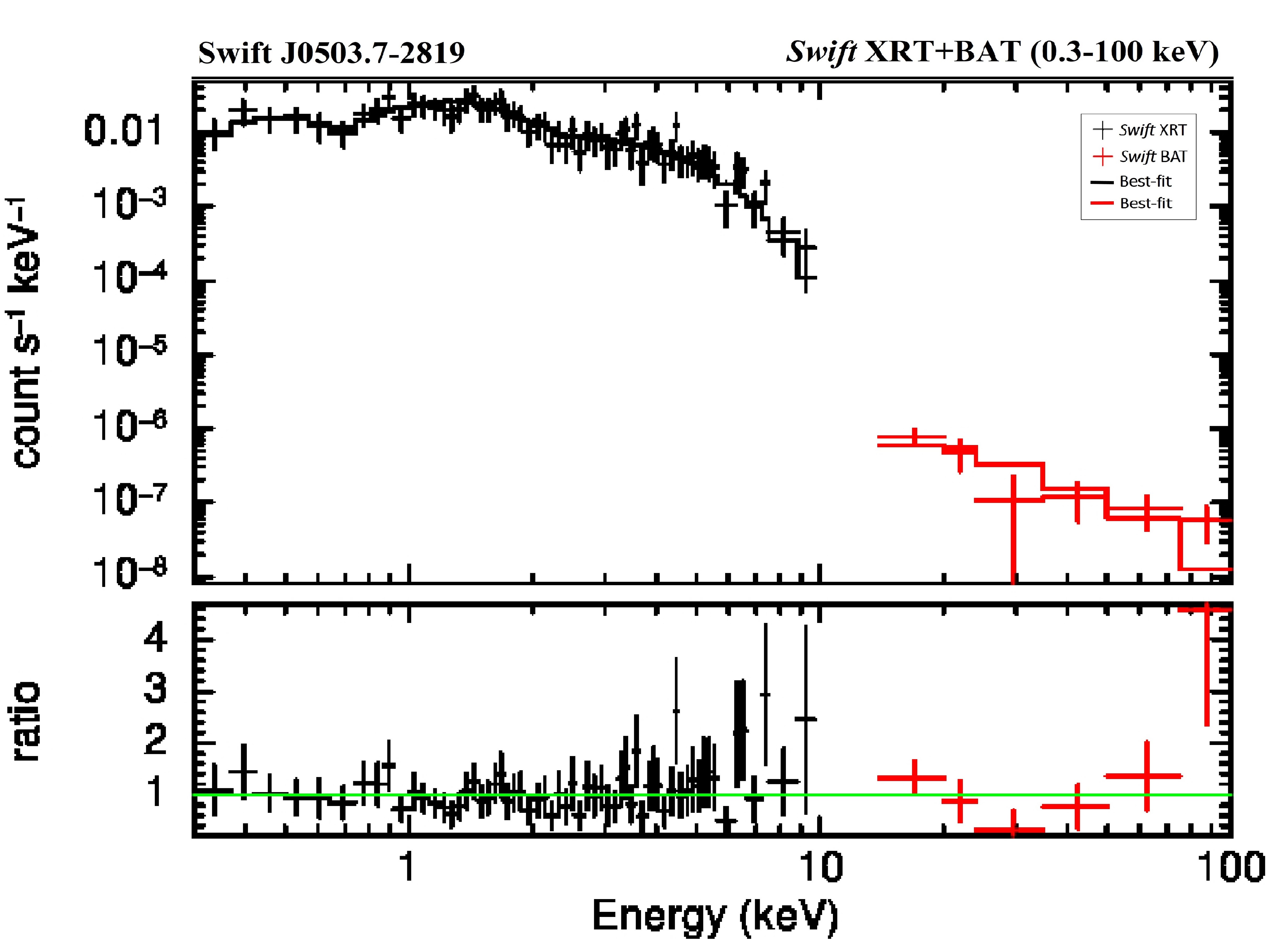}
 \caption{Broad-band 0.3-100 keV X-ray spectrum from {\it Swift} XRT (black) and BAT (red).  The best fit model is shown as a histogram and the residuals from the best fit are shown in the lower panel.}
    \label{fig:XRT+BAT_Spectra}
\end{figure*}
\begin{table*}
\caption{\textit{AstroSat} SXT (0.35-5 keV) and {\it{Swift}} XRT+BAT spectral fit model parameters for \textit{Swift}~J0503.7-2819. The abundance (Z) for all elements is set relative to solar values and the table in use is `aspl’. Subscripts 1 and 2 refer to additive models (bbody or apec) in the order of their appearance in the fitted model. The errors and upper limits on the parameters were obtained with the criterion of $\chi^{2} _{min} + 2.73$ for 90 \% confidence}
\label{tab:Spectral_Fit_Table}
\resizebox{2\columnwidth}{!}{
\begin{tabular}{|c|ccccccccccc|}
\hline
\textbf{Spectral Model} &
  \multicolumn{11}{c|}{\textbf{Parameters}} \\\hline
  \textbf{SXT- 1+2} & & & & & & & & & & &\\
\hline

\textbf{} &
  \multicolumn{1}{c|}{\textbf{N$_{H_{tbabs}}$}} &
   \multicolumn{1}{c|}{\textbf{N$_{H_{pcfabs}}$}} &
  \multicolumn{1}{c|}{\textbf{pcf}} &
  \multicolumn{1}{c|}{\textbf{kT$_{1}$ (bbody/apec)}} &
  \multicolumn{1}{c|}{\textbf{A$_{1}$}} &
  \multicolumn{1}{c|}{\textbf{kT$_{2}$ (apec)}} &
  \multicolumn{1}{c|}{\textbf{A$_{2}$}} &
  \multicolumn{1}{c|}{\textbf{$\chi$$_{\nu}$$^{2}$/dof}} &
  \textbf{Flux (0.3-2.0) keV)} &
  \textbf {Flux (2-10) keV)} & \\ &
  \multicolumn{1}{c|}{10$^{20}$ (cm$^{-2}$)} &
   \multicolumn{1}{c|}{10$^{20}$ (cm$^{-2}$)} &
 \multicolumn{1}{c|}{\%} &
  \multicolumn{1}{c|}{(keV)} &
  \multicolumn{1}{c|}{ph cm$^{-2}s^{-1}$} &
  \multicolumn{1}{c|}{(keV)} &
  \multicolumn{1}{c|}{ph cm$^{-2}s^{-1}$} &
  \multicolumn{1}{c|}{} &
  erg s$^{-1}$ cm$^{-2}$ & erg s$^{-1}$ cm$^{-2}$ &
 \\ \hline
tbabs*bbody &	
\multicolumn{1}{c|}{1.08} &
\multicolumn{1}{c|}{...} &
\multicolumn{1}{c|}{...} &
  \multicolumn{1}{c|}{0.78} &
  \multicolumn{1}{c|}{2.54×10$^{-5}$} &
  \multicolumn{1}{c|}{...} &
  \multicolumn{1}{c|}{...} &
  \multicolumn{1}{c|}{2.8/59} &
  0.67$\times$10$^{-12}$ & 1.5$\times$10$^{-12}$ & \\ \hline
tbabs*apec &
\multicolumn{1}{c|}{1.08} &
\multicolumn{1}{c|}{...} &
\multicolumn{1}{c|}{...} &
  \multicolumn{1}{c|}{...} &
  \multicolumn{1}{c|}{...} &
  \multicolumn{1}{c|}{$>$60} &
  \multicolumn{1}{c|}{2.57×10$^{-3}$} &
  \multicolumn{1}{c|}{1.6/59}&
  1 $\times$10$^{-12}$ & 3.0$\times$10$^{-12}$ & \\ \hline
  tbabs*(apec+apec) &
\multicolumn{1}{c|}{1.08} &
\multicolumn{1}{c|}{...} &
\multicolumn{1}{c|}{...} &
  \multicolumn{1}{c|}{0.076$^{+0.030}_{-0.031}$} &
  \multicolumn{1}{c|}{6.9$\times$10$^{-3}$} &
  \multicolumn{1}{c|}{$>$60} &
  \multicolumn{1}{c|}{2.38×10$^{-3}$} &
  \multicolumn{1}{c|}{1.18/56}&
   1.81$\times$10$^{-12}$ & 2.97$\times$10$^{-12}$ & \\ \hline
   tbabs*(bbody+apec) &
\multicolumn{1}{c|}{1.08} &
\multicolumn{1}{c|}{...} &
\multicolumn{1}{c|}{...} &
  \multicolumn{1}{c|}{0.064$^{+0.023}_{-0.026}$} &
  \multicolumn{1}{c|}{2.61$\times$10$^{5}$} &
  \multicolumn{1}{c|}{$>$50} &
  \multicolumn{1}{c|}{2.41×10$^{-5}$} &
  \multicolumn{1}{c|}{1.18/58}&
  1.49$\times$10$^{-12}$ & 2.95$\times$10$^{-12}$ & \\ \hline
  tbabs*pcfabs(bbody+apec) &
\multicolumn{1}{c|}{1.08} &
\multicolumn{1}{c|}{19.5} &
\multicolumn{1}{c|}{63.5} &
  \multicolumn{1}{c|}{0.064$^{+0.029}_{-0.025}$} &
  \multicolumn{1}{c|}{8.24$\times$10$^{-5}$} &
  \multicolumn{1}{c|}{33.4$^{+24.15}_{-33.76}$} &
  \multicolumn{1}{c|}{5.8×10$^{-3}$} &
  \multicolumn{1}{c|}{1.05/58}&
   1.46$\times$10$^{-12}$ & 5.17$\times$10$^{-12}$ & \\ \hline  
  \textbf{Swift XRT+BAT} & & & & & & & & & & &\\
\hline
tbabs*(apec+apec) &
\multicolumn{1}{c|}{27$^{+16}_{-10}$} &
\multicolumn{1}{c|}{...} &
\multicolumn{1}{c|}{...} &
  \multicolumn{1}{c|}{0.050$^{+0.034}_{-0.018}$} &
  \multicolumn{1}{c|}{0.29} &
  \multicolumn{1}{c|}{$>$42} &
  \multicolumn{1}{c|}{1.63x10$^{-3}$} &
  \multicolumn{1}{c|}{0.62/313}&
  0.5$\times$10$^{-12}$ & 2.0$\times$10$^{-12}$ & \\ \hline
\end{tabular}
}
\end{table*}
The BAT spectrum of Swift~J0503.7-2819 was obtained from the BAT 70 Month catalogue\footnote{\url{https://swift.gsfc.nasa.gov/results/bs70mon/SWIFT_J0503.7-2819}}, and a joint fit is performed to the BAT and XRT spectra over 0.3--100~keV, Figure~\ref{fig:XRT+BAT_Spectra}. A two-temperature fit is preferred at $>$3$\sigma$ (Table~\ref{tab:Spectral_Fit_Table}), although only a lower limit can be placed on the temperature of the hotter component, consistent with the \textit{AstroSat} results.

\subsubsection{{\it{SALT Spectra}}}\label{subsec:SALT_Spectra}
\begin{figure*}
\includegraphics[width=0.96\textwidth]{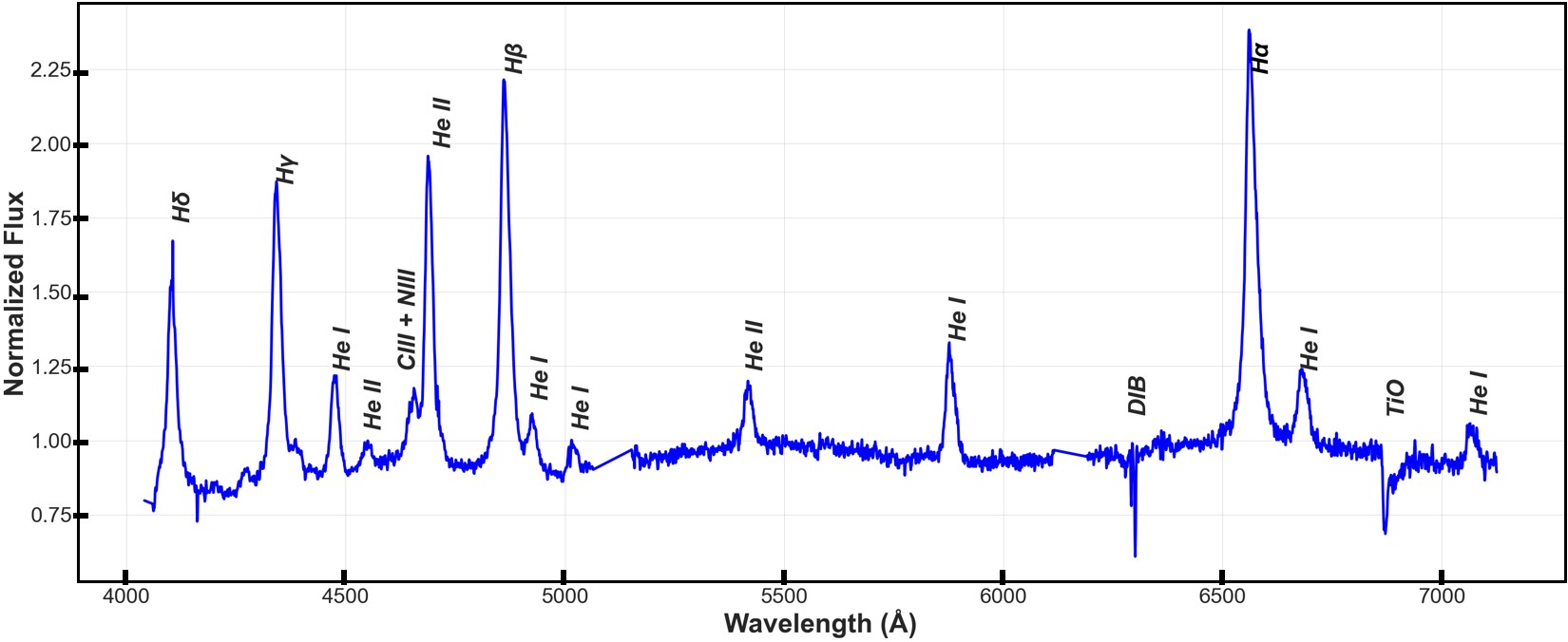}
\caption{The mean optical spectrum of Swift~J0503.7-2819.  Diffuse interstellar bands (DIB), Telluric absorption features are not removed.}
\label{Fig:mean_spectrum}
\end{figure*}
\begin{figure*}
\includegraphics[width=\columnwidth]{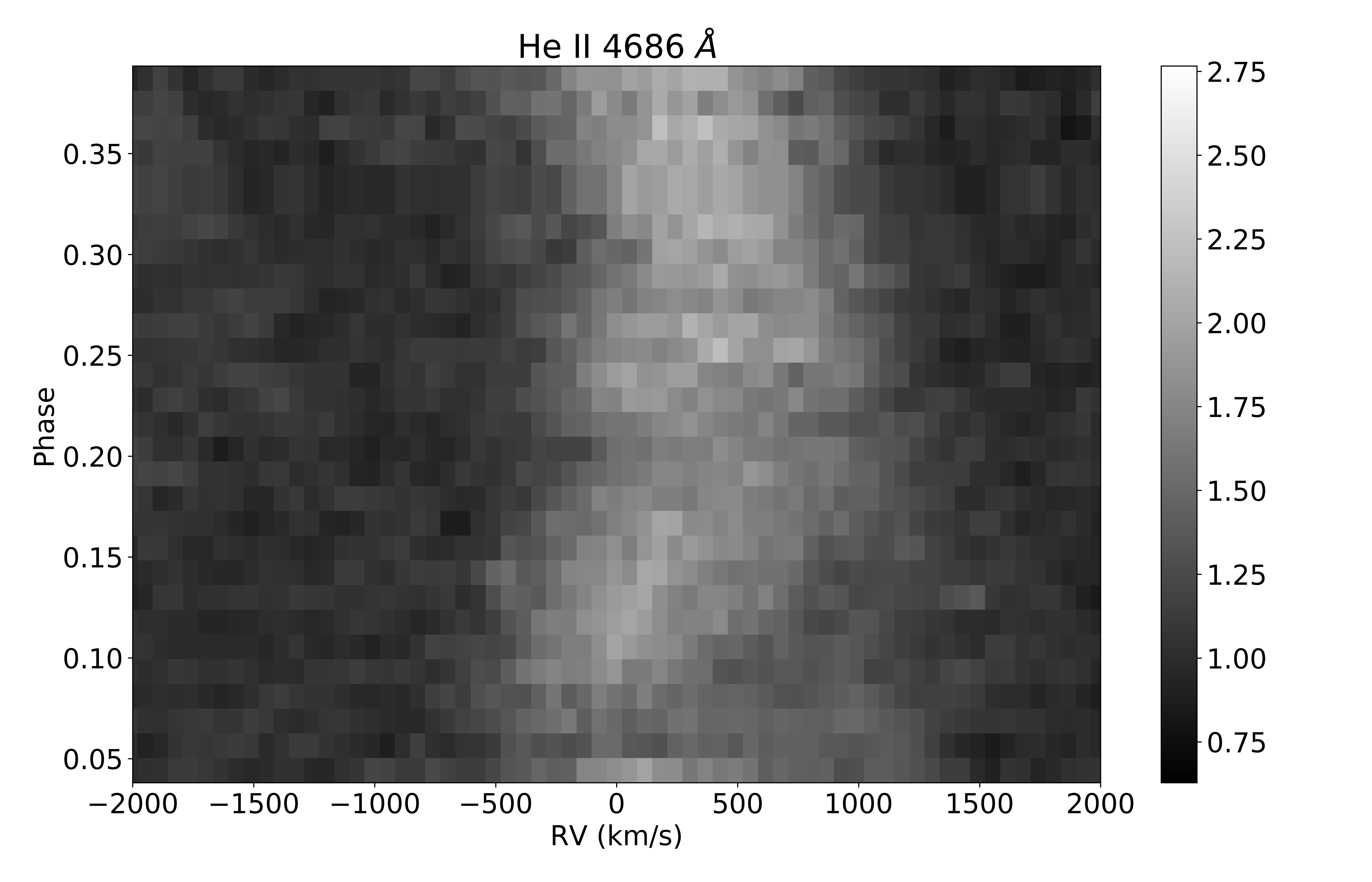}
\includegraphics[width=\columnwidth]{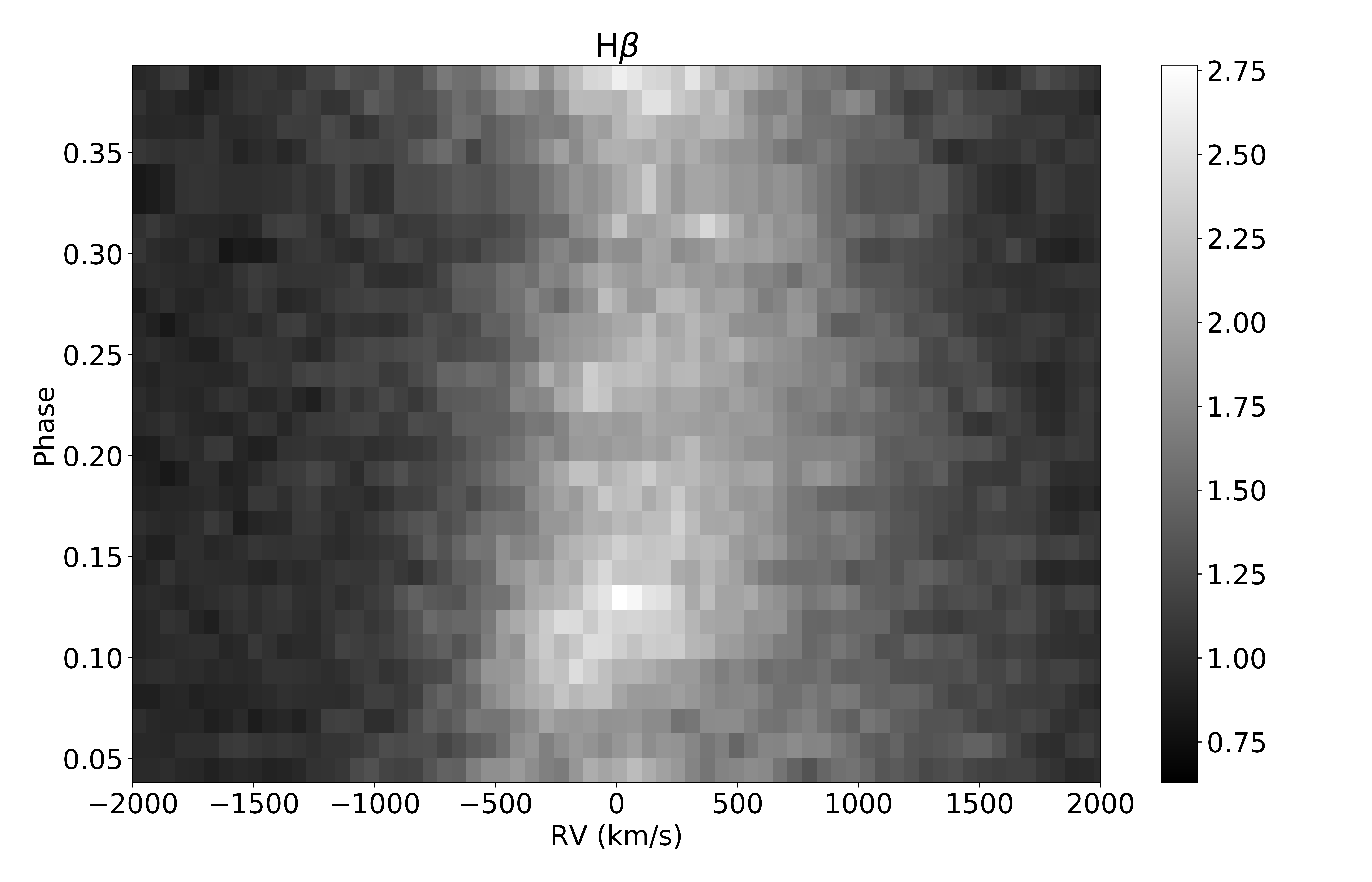}
\caption{2D dynamic spectra of Swift~0503.7-2819 phased to the orbital period, centered at He II 4686\,\AA\, (left) and at H$\beta$ (right).}
\label{Fig:dynamic_spec_Orb}
\end{figure*}

\begin{figure*}
\includegraphics[width=\textwidth]{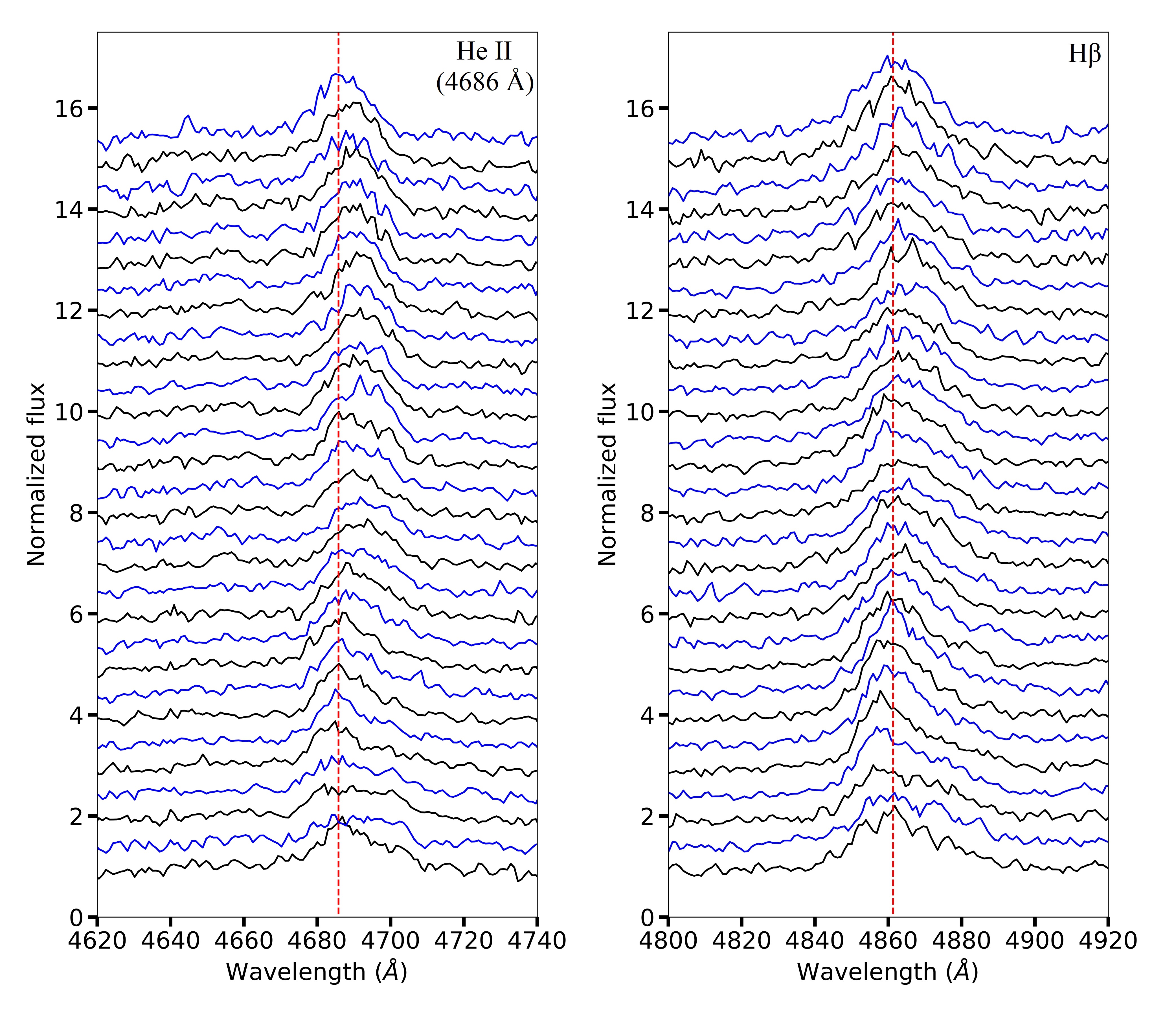}
\caption{The line profiles evolution of He II 4686\,\AA\ (left), and H$\beta$ (right) plotted in chronological order from bottom to top. The red dashed lines represent the rest wavelength of the emission lines. We add an incremental vertical offset of 0.5 to each of the normalized spectra for visualization purposes.}
\label{Fig:line_profiles_HeII_V2}
\end{figure*}

\begin{figure*}
\includegraphics[width=0.5\textwidth]{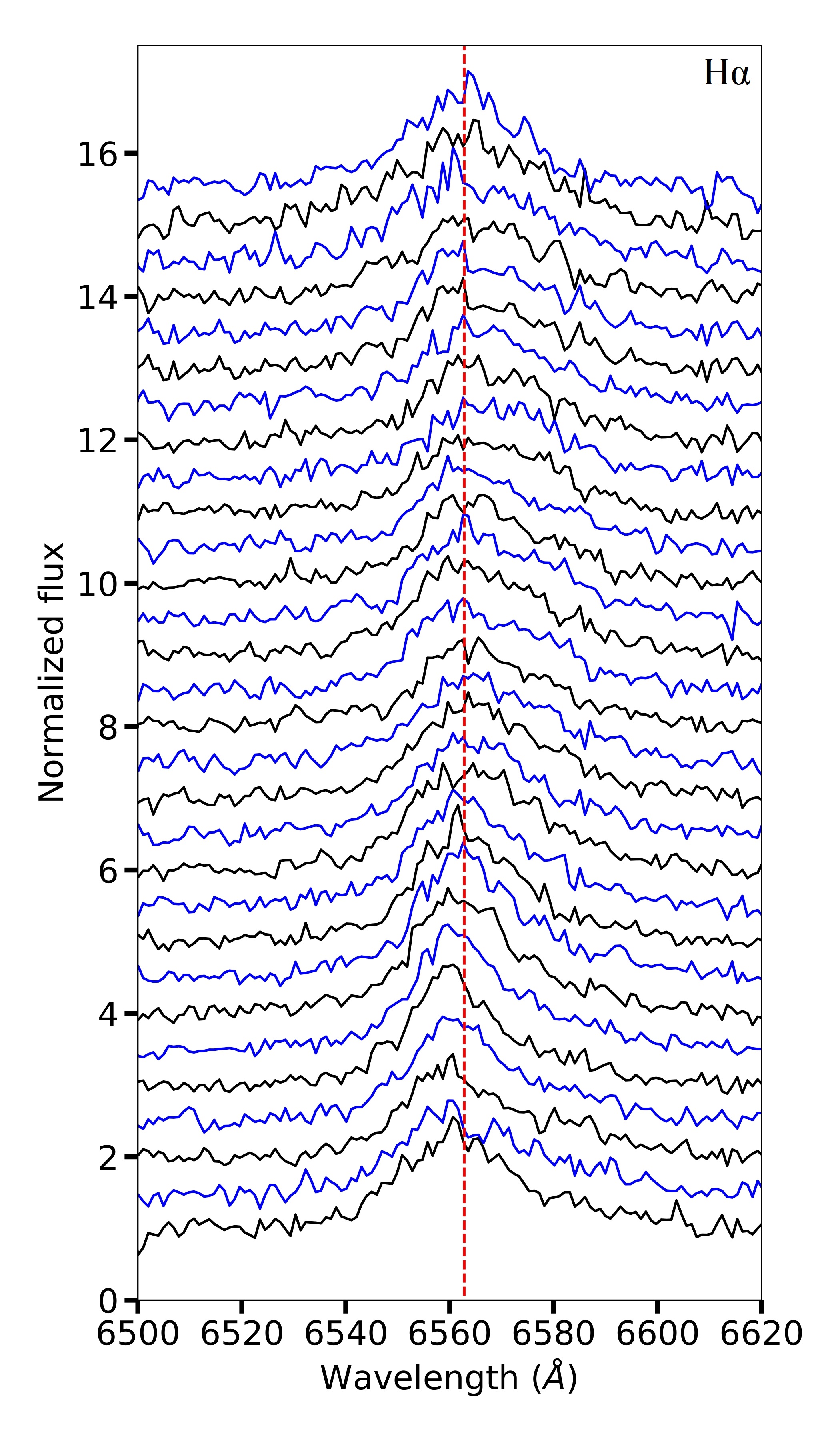}
\caption{The line profiles evolution of H$\alpha$ plotted in chronological order from bottom to top. The red dashed lines represent the rest wavelength of H$\alpha$. We add an incremental vertical offset of 0.5 to each of the normalized spectra for visualization purposes.}
\label{Fig:line_profiles_Halpha_V2}
\end{figure*}
\begin{figure*}
\includegraphics[width=\textwidth]{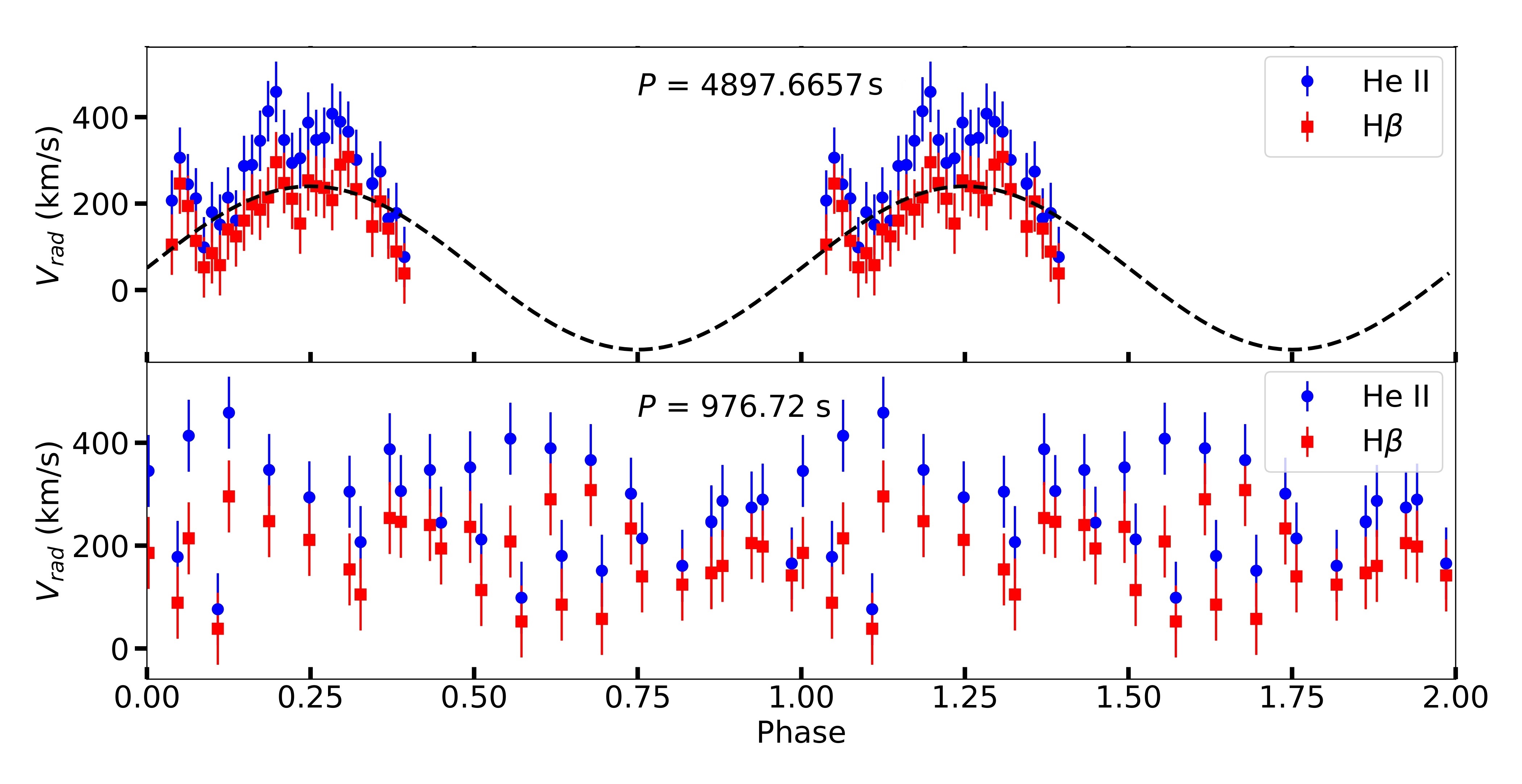}
\caption{The radial velocities measured from single Gaussian fitting to the H$\beta$ and He II 4686\,\AA, phased over the $P_{orb}$ and the QPO. The fitted sinusoid from \protect\cite{2015AJ....150..170H} is overlaid on the orbit-folded velocity curve. Phase zero corresponds to BJD 2456683.6274. A barycentric correction is applied to the radial velocities.}
\label{Fig:phased_RVs}
\end{figure*}
The mean spectrum obtained from observations with the SALT is shown in Figure~\ref{Fig:mean_spectrum}, with the optical lines marked. The total length of the observation is 1920 s, covering $\sim$ 40\% of the orbital cycle. In Figures~\ref{Fig:dynamic_spec_Orb} we present the RSS time-series spectroscopy as 2-D dynamic spectra centered on He II 4686\,\AA\ and H$\beta$ phased to the orbital period. There is a clear change in the radial velocities towards a redshift in the middle of the time series, this is again evident in Figures \ref{Fig:line_profiles_HeII_V2} and \ref{Fig:line_profiles_Halpha_V2}. The lines show significant changes with time and likely consist of multiple components. This is expected from magnetic CVs where emission lines show complex structures associated with emission from the accretion stream, the flow through the magnetosphere, and the secondary star \cite[e.g.,][]{1987PhDT.......130R,1995cvs..book.....W,1997A&A...319..894S,2018MNRAS.480..572A}. The redshifting of the radial velocity is likely due to the orbital modulation as reported by \cite{2022ApJ...934..123H}, or the spin period if the optical light arises from material coupled to the WD magnetosphere and so also the spin.  

We measured the radial velocities of both H$\beta$ and He II 4686\,\AA\ and we phase fold them over multiple periods (see Figure~\ref{Fig:phased_RVs}). The radial velocities are measured by plotting a single Gaussian to the line profiles. Since emission lines consist of multiple components, it is not ideal to fit only a single Gaussian to the line profiles, but the resolution of the spectra does not allow to disentangle the different components, therefore we fit the line profiles with a simple, single Gaussian component. The radial velocities are folded on different periods using the ephemeris from \cite{2022ApJ...934..123H}, where phase zero corresponds to the blue-to-red crossing of the emission-line radial velocity, at BJD 2456683.6274(9). Further, the fitted sinusoid from \cite{2015AJ....150..170H} overlaid on the orbital radial velocity curve clearly demonstrates that the phasing is consistent.
 Given that the time-coverage of the observations is shorter than a single orbital/spin phase,  it is not possible to discern the systemic velocity of Swift~J0503.7-2819, nor when exactly the blue-to-red crossing of the emission-lines radial velocity takes place.



\section{Discussion}\label{sec:Discussion}

We have carried out FUV and X-ray timing analysis and optical plus X-ray spectral analysis of the MCV Swift~J0503.7-2819. 
The $P_{\Omega}$ and $P_{\omega}$ values are identified to be 4897.6657 s and  3932.0355 s, respectively, and are commensurate with the values reported by \cite{Rawat:2022onw}, and model 1 of \cite{2022ApJ...934..123H}. All orbital light curves are aligned in phase using the ephemeris from \cite{2022ApJ...934..123H}, with phase 0 corresponding to the epoch of blue-to-red crossing of the emission line radial velocity.  

The signal at 0.20499 mHz in the extremely noisy FUV power spectrum is identified with the orbital period of the system. The orbital FUV light curve shows a double-humped structure. The light curve profile can be subjected to many interpretations but as noted by \cite{2022ApJ...934..123H} in the case of the optical light curve, the most plausible explanation for orbital signatures in the light curve point to its contribution by asymmetric structures of a disc.

The X-ray power spectra show four conspicuous peaks; they are, in the order of strengths $\omega$, $\omega$-$\Omega$, $\Omega$, and $\omega$+$\Omega$. The presence of orbital frequency and the sidebands is typically judged to be the hallmark of disc-less accretion. However, categorizing the accretion flow in atypical systems solely based on the power spectra can be misleading. Therefore, we look at the possible causes behind individual power spectral signals both in cases of disc-less and disc-ed accretion and examine which better explains the observations in Swift~J0503.7-2819. 

The spin modulation seen in the light curves of MCVs can result from the self-occultation of the emitting regions by the rotating WD body and/or the photo-electric absorption/electron-scattering of the emitted radiation by infalling material. The energy-dependent spin-pulsed light curve of Swift~J0503.7-2819 can only be explained by taking into account both the effects of occultation and phase-varying absorption. The `accretion curtain' scenario \citep{1988MNRAS.231..549R} with an arc-shaped accretion structure around the WD can adequately explain the spin modulation seen in Swift~J0503.7-2819. 

The outer regions of a disc (if present) can retain non-axisymmetric structures which when crossing the line of sight to the WD can reduce the X-ray flux causing orbital modulation. \cite{1993MNRAS.260..299H} proposed an `Obscuration model' for high-inclination IPs, in which energy-dependent orbital modulation results from accretion structures raised by the impacting stream. The orbital modulation can also arise in disc-less systems if the accretion flow is binary-phase dependent i.e., the accreting matter is transferred directly from the secondary via a stream without circularising  and forming a disc. Alternatively, the direct impact of the stream onto the magnetosphere can give rise to orbitally localized regions at the point of contact \citep{1993MNRAS.260..299H}. The orbital light curve of Swift~J0503.7-2819 shows explicit energy dependence as evidenced by the strong orbital modulation of the SR curve. Therefore, photo-electric absorption plays a major role in producing the modulation and is likely caused by material in the binary frame.  

The disparity in power between the two sidebands shows there is intrinsic modulation at these frequencies \citep{1992MNRAS.255...83W}. A purely disc-fed model does not permit X-ray modulation at any beat period \citep{1991sepa.conf..213K, 1991MNRAS.248P...5H}. The presence of a positive beat signal, $\omega$+$\Omega$, is puzzling. Theoretically, a retrograde motion of the WD would produce a signal at $\omega$+$\Omega$. But this geometry is highly unlikely as there is no obvious reason why the WD would spin against the orbital motion. If the WD is indeed executing a retrograde spin the torque exerted on it by the accreting matter and the drag force due to magnetic stress would cause it to spin-down \citep{1978ApJ...223L..83G, 1983ASSL..101..229L}. Disc-less accretion could also explain the $\omega$+$\Omega$ provided the symmetry between the poles is broken i.e., extreme disparities in properties like size, luminosity, etc. \citep{1992MNRAS.255...83W}.   
The negative beat period modulation present in the X-ray light curve can be explained by the `pole-switching' that is expected to take place every half-beat cycle. As detailed in Section \ref{subsec:X-ray_LCs_power}, a complete pole-switching is assumed absent in Swift~J0503.7-2819, instead accretion stream switches between feeding only the `upper' pole and feeding both poles simultaneously. Magnetohydrodynamical modeling of asynchronous systems with complex magnetic field configurations by \cite{2012ARep...56..257Z} shows that pole-switching can also be accompanied by transitions from One/Two-pole accretions, consistent with the observations of Swift~J0503.7-2819. The presence of $\omega$ - $\Omega$ modulation does not necessarily imply a stream-fed accretion, however. Proposed by \cite{1992MNRAS.258..697N},  a system possessing a `non-accretion' structure \citep{1991sepa.conf..213K,1991ApJ...378..674K} can also produce negative beat period modulation. In this scenario, a truncated non-accretion disc or a ring is present, which preferentially feeds the pole nearer to the inner Lagrangian point. This will switch over to the opposite pole half a beat cycle later. The switching of accretion between poles is expected to be gradual and not instantaneous. In Swift~J0503.7-2819, the intervals showcasing the transition from one-pole to two-pole and vice versa, are marked by erratic activity like flaring and lasts for a sizeable duration of the cycle. The X-ray counterpart to the optical signal at 975 s \citep{2015AJ....150..170H} is likely caused by this X-ray flaring. Though the`non-accretion' structure would mimic an accretion disc in its observational characteristics the accretion flow in the system is not carried out through a disc at all. The systems possessing such a non-accretion structure would therefore exhibit properties of both geometries.

The X-ray spectra of Swift~J0503.7-2819 are fitted using a two-temperature model, the temperatures of the cooler component in the soft X-ray {\it{AstroSat}}-SXT spectra and broad-band {\it{Swift}} XRT+BAT spectra are comparable, $\sim64$ eV and $ \sim50$ eV, respectively. The temperature of the hotter component in the soft X-ray spectra ($\sim$33.4 keV) could only be constrained by introducing a partial absorber with a hydrogen column density equivalent of $\sim 2.7 \times 10^{21} cm^{-2}$ and a covering fraction of 63.5\%. 

The optical spectra of Swift~J0503.7-2819 show broad, symmetric emission lines on top of an intrinsically flat or blue continuum, a characteristic typical of disc-accreting systems. Unlike polars, which exhibit narrow and asymmetric spectral lines due to the absence of a disc, disc-ed systems resemble non-magnetic CVs in their line profiles, i.e., large widths and often double peaks. 
The red-shifting of the radial velocity curve is likely due to the orbital modulation of the emission lines, however, it is hard to ascertain the source of the spectroscopic period as explained by \cite{2022ApJ...934..123H}. It is postulated that in disc-ed systems the optically thick H~I lines have their origin in the outer regions of the disc while the He~II line is likely produced in a localized region like the bright spot \citep{1980ApJ...235..939W}, in this case, the emission lines are expected to show orbital modulation. The equivalent width (EW) ratio of H$\alpha$/ H$\beta$ in Swift~J0503.7-2819 is 1.27, an EW ratio falling between 1 and 2.86 is suggestive of significant photon-trapping effects \citep[see][and references therein]{2019Ap&SS.364..153M}. Photon trapping effects are common in scenarios involving optically thick accretion flows such as those found in accretion discs/curtains. Accretion structures facilitate the interaction between matter and photons, delaying the liberation of the latter  emanating from deep within \citep{2002ApJ...574..315O}. Further, IPs can be characterized by a H$\beta$ EW greater than 20\,\AA \footnote{EWs of emission lines are expressed in positive values} and He~II/H$\beta$ EW ratio greater than 0.4 \citep{1992PhDT.......119S}. The EW of H$\beta$ in Swift~J0503.7-2819 is estimated to be 39.35 \,\AA\ and He~II/H$\beta$ EW ratio is 0.73. While Swift~J0503.7-2819 satisfies these criteria, they are not sufficient on their own to classify the system as an IP.
The foregoing observations of Swift~J0503.7-2819 show that the system does not conform to a purely disc-fed or stream-fed model. The optical characteristics of the system and the significant photoelectric absorption suffered by the X-ray emissions suggest the presence of accretion structures around the WD. However, the high $P_{\omega}/P_{\Omega}$ ratio of the system would mean that the co-rotation radius ($R_{co}$) is much greater than the circularization radius ($R_{cir}$), therefore it is not possible for the system to possess a Keplerian disc. In the following section, the nature of the accretion flow in Swift~J0503.7-2819 is explored. 

\subsection{The Nature of the Asynchronicity and Accretion flow in Swift~J0503.7-2819} \label{subsec:flow_spin}

Swift~J0503.7-2819 has a $P_{\omega}$/$P_{\Omega}$ ratio of 0.8 and asynchronicity (1-$P_{\omega}$/$P_{\Omega}$) of 19.7$\%$. With $P_{\omega}$/$P_{\Omega}$ > 0.1 and an orbital period below the period gap, Swift~J0503.7-2819 falls into the category of EX Hya-like systems. \cite{1999MNRAS.310..203K} suggested the high $P_{\omega}$/$P_{\Omega}$ of EX Hya is indicative of a new spin equilibrium different from that followed by conventional IPs, and proposed the following criterion for stability,
\begin{equation} 
R_{co} \sim b \\.
\end{equation}
\noindent where `b' is the distance from the WD to the inner Lagrangian point.
That is, the spin can be understood as an equilibrium condition in which the magnetospheric radius of the WD extends out to the inner Lagrangian point.  Further, any system occupying this equilibrium should also satisfy the condition \citep{2004RMxAC..20..138N, 1999MNRAS.310..203K}:
  
\begin{equation}\label{equ:Equation2}
    \frac{P_{\omega}}{P_{\Omega}} \sim(0.500-0.227 \log q)^{3 / 2},
\end{equation}

\noindent where `q' is the mass ratio.
Based on the newly observed spin equilibria in EX Hya, \cite{1999MNRAS.310..203K} put forth the argument that there should exist a continuum of spin equilibria appropriate to the magnetic moment and orbital period of a given IP. 

\cite{2004RMxAC..20..138N} explored this parameter space and investigated the equilibria using a magnetic accretion model. They found the identified spin equilibria correspond to different types of accretion flows vis-{\` a}-vis,
at small values of $P_{\omega}$/$P_{\Omega}$ ($\leq 0.1$) the accretion is disc-like, at intermediate values ($0.1\le$ $P_{\omega}$/$P_{\Omega}$ $\le 0.6$), it is streamlike. At higher values of $P_{\omega}$/$P_{\Omega}$ like in case of EX Hya the accretion is fed from a ring at the outer edge of the WD Roche lobe. \cite{2004ApJ...614..349N} proposed  the growing number of EX Hya-like systems with high $P_{\omega}$/$P_{\Omega}$ are likely to have ring-fed accretion. Further, it is apparent that as IPs evolve to shorter periods (below the period gap) and smaller mass ratios, the $P_{\omega}$/$P_{\Omega}$ generally increase to maintain the spin equilibrium. Those that are unable to synchronize will appear similar to EX Hya with a large period ratio and ring-like accretion flow \citep{2007MNRAS.378..211M, 2008ApJ...672..524N}.
If Swift~J0503.7-2819 occupies a similar spin equilibrium as shown by EX Hya, the observed period ratio should be commensurate with the result of Equation \ref{equ:Equation2} within reasonable limits. Using the mean empirical mass-period relation given by \cite{1998MNRAS.301..767S} the secondary star has an estimated mass of 0.085 $M_\odot$. Using the WD mass determined by \cite{Rawat:2022onw} of 0.54$M_\odot$, the mass ratio for Swift~J0503.7-2819 becomes, q = 0.159. \cite{2004ApJ...614..349N} have constructed their model for a q value of 0.5. Using the q value appropriate for Swift~J0503.7-2819 i.e., q = 0.159, the equilibrium condition becomes:  

\begin{equation}
\left(\frac{P_{\omega}}{P_{\Omega}}\right)_{q=0.159} \simeq 1.3126\left(\frac{P_{\omega}}{P_{\mathrm{\Omega}}}\right)_{q=0.5} ,
\end{equation}
From \cite{2004RMxAC..20..138N},

\begin{equation}
\left(\frac{P_{\omega}}{P_{\Omega}}\right)_{q=0.5} \sim 0.54 .
\end{equation}

\noindent This predicts a $P_{\omega}$/$P_{\Omega}$ of $\approx$ 0.71 for Swift~J0503.7-2819. 
The slightly higher ratio observed might be indicative that instead of forming right on the edge of the WD lobe, the ring structure occurs just outside of it \citep{2004ApJ...614..349N}\footnote{This is only a heuristic approach, the (Hydisc code) model used by \cite{2004ApJ...614..349N} is also contingent on the coefficient `k' which embodies the plasma-magnetic field interaction.}. \cite{2004RMxAC..20..138N} speculate that EX Hya-like systems are unable to synchronize due to the very low magnetic moments of the associated secondaries. From \cite{1995cvs..book.....W} the surface magnetic moment of the secondary can be approximated as,
 \begin{equation} \label{Equ:MagMoment}
\mu_2=2.8 \times 10^{33} P_{\mathrm{\Omega}}^{9 / 4} \mathrm{G} \mathrm{cm}^3
\end{equation} 
If the magnetic moment of the secondary is lower than what is predicted by this value, the system will appear like EX Hya instead of synchronizing, i.e., evolving into a polar. For Swift~J0503.7-2819, Equation \ref{Equ:MagMoment} yields a value of 5.5$\times$ $10^{33}$ $Gcm^{3}$. The secondary star in the system (either an M dwarf or an early brown-dwarf, from mass estimate) probably has a magnetic moment less than this estimate, hence the asynchronism.

\cite{2014MNRAS.442.2580P} observe that though the hard X-ray luminosity of most IPs is at about $10^{33}$ erg s$^{-1}$ or higher, it seems there is an entirely separate group constituted by low-luminosity IPs (LLIPs). Mukai\footnote{https://asd.gsfc.nasa.gov/Koji.Mukai/iphome/catalog/llip.html} has consolidated both confirmed or ironclad IPs with X-ray luminosity below $2.5\times10^{32}$ erg s$^{-1}$ into LLIPs. There is a striking correlation between LLIPs and EX Hya-like systems. Among the 13 listed LLIPs: nine are IPs within or below the period gap, of which, five namely, EX Hya \citep{1998MNRAS.295..167A}, DW Cnc \citep{2004PASP..116..516P}, HT Cam \citep{2002PASP..114..623K}, V1025 Cen \citep{2002MNRAS.333...84H}, and V598 Pegasi \citep{2007MNRAS.378..635S} come under EX Hya-like systems. The four, namely, CC Scl \citep{2012MNRAS.427.1004W}, AX J1853.3-0128 \citep{2013AJ....146..107T}, CTCV J2056-3014 \citep{2020ApJ...898L..40L}, and V455 And \citep{2005A&A...430..629A}, have a $P_{\omega}/P_{\Omega}$ $<$ 0.1. The four IPs above the period gap include DQ Her \citep{1995ApJ...454..447Z} the LLIP nature of which is ambiguous, a propeller and probable propeller system viz., AE Aqr \citep{1994MNRAS.267..577D}, and V1460 Her \citep{2020MNRAS.499..149A}, respectively and DO Dra \citep{1997ApJ...476..847H} with $P_{\omega}/P_{\Omega}$ $<<$ 0.1.  The low luminosities seen in confirmed LLIPs are attributed to the lower shock temperatures associated with taller shock heights as well as their low accretion rates. Most LLIPs are short-period systems below the period gap with an average rate of mass transfer of 10$^{-10}$ solar masses per year, at the most \citep{2011ApJS..194...28K}.  With a BAT luminosity of $2.4 \times 10^{32}$ erg s$^{-1}$ \footnote{https://swift.gsfc.nasa.gov/results/bs157mon/} and an accretion rate of $1.44 \times 10^{-10} M_\odot yr^{-1}$ \citep{Rawat:2022onw}, Swift~J0503.7-2819 is identified to be an LLIP.

The classification of Swift~J0503.7-2819 is a matter of definition. Asynchronous polars are defined as MCVs temporarily dislodged from their synchronism (due to nova eruptions), the difference between the orbital and spin period in prototypical APs is 2\% or less. A nova eruption, while powerful, is most likely insufficient to produce the extent of asynchronism seen in Swift~J0503.7-2819. For this reason, and others discussed in \cite{2023ApJ...943L..24L}, assigning AP nature to the system is not preferred. Moreover, APs are distinguishable from EX Hya-like systems in lacking spin equilibria, which drives them towards synchronism more rapidly.  The stability criteria and the accretion mode it reveals for Swift~J0503.7-2819 resembles those of the growing class of nearly synchronous MCVs which exhibit characteristics betwixt and between polars and IPs. In attributing AP nature to Swift~J0503.7-2819 \cite{2022ApJ...934..123H} also drew a parallel between the system and the asynchronous Paloma \citep{2007A&A...473..511S, 2016ApJ...830...56J}. Paloma has an asynchronicity of 16.6 \% and  $P_{\omega}/P_{\Omega}$ $\sim$ 0.87, similar to Swift~J0503.7-2819. The orbital period of Paloma falls in the period gap ($P_{\Omega}$ = 2.6 hrs). However, Paloma is judged to most likely be a nearly synchronous IP in Mukai’s IP catalogue, or according to \cite{2016ApJ...830...56J}, an intermediate system between the two classes. From the X-ray luminosity of Paloma \citep{2016ApJ...830...56J}, it is seen to be much like Swift~J0503.7-2819 and EX Hya, also an LLIP, and appears within the period gap with $P_{\Omega}$ of $<$ 3 hr, hence also an EX Hya like system. Swift~J0503.7-2819 also carries a strong resemblance to the SDSS J134441.83+204408.3 ($P_{\Omega}$ =6840 s or 1.9 hrs, and $P_{\omega}/P_{\Omega}$ $\sim$ 0.893), which was reclassified from a synchronous polar to an `asynchronous MCV' by \cite{2023ApJ...943L..24L} earlier this year.

\section{Summary and Conclusions}\label{sec:Summary}
Our multi-wavelength temporal and spectral study of the asynchronous MCV, Swift~J0503.7-2819 reveals the following characteristics:

\begin{enumerate}
    \item X-ray and FUV power spectral estimates of the spin and orbital period of the system show the system is slightly away from synchronism, with an orbital period below the period gap of CVs. 
    \item Optical characteristics of the system and the significant photoelectric absorption suffered by the X-ray emissions point to the presence of accretion structures around the WD. 
    \item Based on numerical simulations exploring similar systems, the high $P_{\omega}$/$P_{\Omega}$ value suggests the accretion flow in Swift~J0503.7-2819 is ring-like, with a non-Keplerian ring feeding the accretion stream. 
    \item The extraordinarily high asynchronicity of 19.7\% 
    cannot be accounted for by a nova eruption which underlies the asynchronicity seen in APs. We propose that unlike APs, and similar to the growing number of asynchronous MCVs which fall into the category of EX Hya-like systems, Swift~J0503.7-2819 is in rotational equilibrium and the asynchronism is stable.
    \item The secondary star (either an M-dwarf or an early brown dwarf from mass estimate) likely possesses a weak magnetic moment preventing spin-orbit synchrony and the system's evolution into a polar.
    \item The system shares many similarities with the growing number of asynchronous `EX Hya-like' MCVs: high $P_{\omega}$/$P_{\Omega}$, an accretion flow not conforming to a purely disc-fed/stream-fed model, low X-ray luminosity and other chimeric characteristics that blur the line between polars and IPs.  
    
    \end{enumerate}
\vspace{-0.3cm}
\section*{Acknowledgements}
We express our sincere gratitude to Prof. Jules Halpern, the referee, for his insightful comments and suggestions. His constructive feedback has significantly enhanced the quality of our work, including improved period determinations using ATLAS data. We thank the Indian Space Research Organisation for scheduling the observations within a short period of time and the Indian Space Science Data Centre (ISSDC) for making the data available.  Kulinder Pal Singh
thanks the Indian National Science Academy for support under the INSA Senior Scientist Programme. Elias Aydi acknowledges support by NASA through the NASA Hubble Fellowship grant HST-HF2-51501.001-A awarded by the Space Telescope Science Institute, which is operated by the Association of Universities for Research in Astronomy, Inc., for NASA, under contract NAS5-26555. K.L. Page acknowledges funding from the UK Space Agency.
This work has been performed utilizing the calibration databases and auxiliary analysis tools developed, maintained, and distributed by {\it AstroSat}-SXT team with members from various institutions in India and abroad and the  SXT Payload Operation Center (POC) at the TIFR, Mumbai for the pipeline reduction. 
A part of this work is based on observations made with the Southern African Large Telescope (SALT), with the Large Science Programme on transients 2018-2-LSP-001 (PI: DAHB)
The work has also made use of software, and/or web tools obtained from NASA's High Energy Astrophysics Science Archive Research Center (HEASARC), a service of the Goddard Space Flight Center and the Smithsonian Astrophysical Observatory. 

\vspace{-0.3cm}
\section*{Facilities:} \textit{AstroSat, ATLAS, SALT, Swift, TESS, XMM-Newton}

\vspace{-0.3cm}
\section*{Data Availability:}
The \textit{AstroSat} data are available in ISSDC at https://astrobrowse.issdc.gov.in/astroarchive/archive/Home.jsp. \\
The \textit{ATLAS} data are available in the ATLAS archive at https://fallingstar-data.com/forcedphot/queue/. \\
The final product of the SALT spectra can be found via this link: https://www.dropbox.com/s/60lcxkjl0flb1w6/Swift\_J0503.zip?dl=0. \\
The \textit{Swift} data used here are available in the \textit{Swift} archives at https://www.swift.ac.uk/swift\_live/, https://heasarc.gsfc.nasa.gov/cgi-bin/W3Browse/swift.pl, and https://www.ssdc.asi.it/mmia/index.php?mission=swiftmastr. \\
The \textit{TESS} data are available in the Barbara A. Mikulski Archive for Space Telescopes (MAST) archives at https://mast.stsci.edu/portal/Mashup/Clients/Mast/Portal.html. \\
The \textit{XMM-Newton} data used here are available at https://heasarc.gsfc.nasa.gov/cgi-bin/W3Browse/w3browse.pl.

\section*{ORCID \MakeLowercase{i}D\MakeLowercase{s}}
Kala G Pradeep \orcidlink{0000-0003-1905-2962}{ https://orcid.org/0000-0003-1905-2962} 
\\
Kulinder Pal Singh \orcidlink{0000-0001-6952-3887}{ https://orcid.org/0000-0001-6952-3887}
\\
G. C. Dewangan \orcidlink{0000-0003-1589-2075}{ https://orcid.org/0000-0003-1589-2075}
\\
Elias Aydi \orcidlink{0000-0001-8525-3442}{ https://orcid.org/0000-0001-8525-3442}
\\
P. E. Barrett \orcidlink{0000-0002-8456-1424}{ https://orcid.org/0000-0002-8456-1424}
\\
D. A. H. Buckley \orcidlink{0000-0002-7004-9956}{ https://orcid.org/0000-0002-7004-9956}
\\
K. L. Page \orcidlink{0000-0001-5624-2613}{ https://orcid.org/0000-0001-5624-2613}
\\
S. B. Potter \orcidlink{0000-0002-5956-2249}{ https://orcid.org/0000-0002-5956-2249}
\\
E. M. Schlegel \orcidlink{0000-0002-4162-8190}{ https://orcid.org/0000-0002-4162-8190}

\bibliographystyle{mnras}
\bibliography{bibliography} 

\begin{thebibliography}{}
\makeatletter
\relax
\def\mn@urlcharsother{\let\do\@makeother \do\$\do\&\do\#\do\^\do\_\do\%\do\~}
\def\mn@doi{\begingroup\mn@urlcharsother \@ifnextchar [ {\mn@doi@}
  {\mn@doi@[]}}
\def\mn@doi@[#1]#2{\def\@tempa{#1}\ifx\@tempa\@empty \href
  {http://dx.doi.org/#2} {doi:#2}\else \href {http://dx.doi.org/#2} {#1}\fi
  \endgroup}
\def\mn@eprint#1#2{\mn@eprint@#1:#2::\@nil}
\def\mn@eprint@arXiv#1{\href {http://arxiv.org/abs/#1} {{\tt arXiv:#1}}}
\def\mn@eprint@dblp#1{\href {http://dblp.uni-trier.de/rec/bibtex/#1.xml}
  {dblp:#1}}
\def\mn@eprint@#1:#2:#3:#4\@nil{\def\@tempa {#1}\def\@tempb {#2}\def\@tempc
  {#3}\ifx \@tempc \@empty \let \@tempc \@tempb \let \@tempb \@tempa \fi \ifx
  \@tempb \@empty \def\@tempb {arXiv}\fi \@ifundefined
  {mn@eprint@\@tempb}{\@tempb:\@tempc}{\expandafter \expandafter \csname
  mn@eprint@\@tempb\endcsname \expandafter{\@tempc}}}

\bibitem[\protect\citeauthoryear{{Allan}, {Hellier}  \& {Beardmore}}{{Allan}
  et~al.}{1998}]{1998MNRAS.295..167A}
{Allan} A.,  {Hellier} C.,   {Beardmore} A.,  1998, \mn@doi [\mnras]
  {10.1046/j.1365-8711.1998.29511353.x}, \href
  {https://ui.adsabs.harvard.edu/abs/1998MNRAS.295..167A} {295, 167}

\bibitem[\protect\citeauthoryear{{Andronov}}{{Andronov}}{1987}]{1987Ap&SS.131..557A}
{Andronov} I.~L.,  1987, \mn@doi [\apss] {10.1007/BF00668138}, \href
  {https://ui.adsabs.harvard.edu/abs/1987Ap&SS.131..557A} {131, 557}

\bibitem[\protect\citeauthoryear{{Araujo-Betancor} et~al.,}{{Araujo-Betancor}
  et~al.}{2005}]{2005A&A...430..629A}
{Araujo-Betancor} S.,  et~al., 2005, \mn@doi [\aap]
  {10.1051/0004-6361:20041736}, \href
  {https://ui.adsabs.harvard.edu/abs/2005A&A...430..629A} {430, 629}

\bibitem[\protect\citeauthoryear{{Ashley} et~al.,}{{Ashley}
  et~al.}{2020}]{2020MNRAS.499..149A}
{Ashley} R.~P.,  et~al., 2020, \mn@doi [\mnras] {10.1093/mnras/staa2676}, \href
  {https://ui.adsabs.harvard.edu/abs/2020MNRAS.499..149A} {499, 149}

\bibitem[\protect\citeauthoryear{{Asplund}, {Grevesse}, {Sauval}  \&
  {Scott}}{{Asplund} et~al.}{2009}]{2009ARA&A..47..481A}
{Asplund} M.,  {Grevesse} N.,  {Sauval} A.~J.,   {Scott} P.,  2009, \mn@doi
  [\araa] {10.1146/annurev.astro.46.060407.145222}, \href
  {https://ui.adsabs.harvard.edu/abs/2009ARA&A..47..481A} {47, 481}

\bibitem[\protect\citeauthoryear{{Aydi} et~al.,}{{Aydi}
  et~al.}{2018}]{2018MNRAS.480..572A}
{Aydi} E.,  et~al., 2018, \mn@doi [\mnras] {10.1093/mnras/sty1759}, \href
  {https://ui.adsabs.harvard.edu/abs/2018MNRAS.480..572A} {480, 572}

\bibitem[\protect\citeauthoryear{{Baumgartner}, {Tueller}, {Markwardt},
  {Skinner}, {Barthelmy}, {Mushotzky}, {Evans}  \& {Gehrels}}{{Baumgartner}
  et~al.}{2013}]{2013ApJS..207...19B}
{Baumgartner} W.~H.,  {Tueller} J.,  {Markwardt} C.~B.,  {Skinner} G.~K.,
  {Barthelmy} S.,  {Mushotzky} R.~F.,  {Evans} P.~A.,   {Gehrels} N.,  2013,
  \mn@doi [\apjs] {10.1088/0067-0049/207/2/19}, \href
  {https://ui.adsabs.harvard.edu/abs/2013ApJS..207...19B} {207, 19}

\bibitem[\protect\citeauthoryear{Bretthorst}{Bretthorst}{1988}]{Bretthorst1988Bayesian}
Bretthorst G.~L.,  1988, Bayesian Spectrum Analysis and Parameter Estimation.
Springer-Verlag Berlin Heidelberg, \url {http://bayes.wustl.edu/glb/book.pdf}

\bibitem[\protect\citeauthoryear{{Buckley}, {Swart}  \& {Meiring}}{{Buckley}
  et~al.}{2006}]{2006SPIE.6267E..0ZB}
{Buckley} D. A.~H.,  {Swart} G.~P.,   {Meiring} J.~G.,  2006, in {Stepp} L.~M.,
   ed.,  Society of Photo-Optical Instrumentation Engineers (SPIE) Conference
  Series Vol. 6267, Society of Photo-Optical Instrumentation Engineers (SPIE)
  Conference Series. p. 62670Z, \mn@doi{10.1117/12.673750}

\bibitem[\protect\citeauthoryear{{Burgh}, {Nordsieck}, {Kobulnicky},
  {Williams}, {O'Donoghue}, {Smith}  \& {Percival}}{{Burgh}
  et~al.}{2003}]{2003SPIE.4841.1463B}
{Burgh} E.~B.,  {Nordsieck} K.~H.,  {Kobulnicky} H.~A.,  {Williams} T.~B.,
  {O'Donoghue} D.,  {Smith} M.~P.,   {Percival} J.~W.,  2003, in {Iye} M.,
  {Moorwood} A. F.~M.,  eds,  Society of Photo-Optical Instrumentation
  Engineers (SPIE) Conference Series Vol. 4841, Instrument Design and
  Performance for Optical/Infrared Ground-based Telescopes. pp 1463--1471,
  \mn@doi{10.1117/12.460312}

\bibitem[\protect\citeauthoryear{{Crawford} et~al.,}{{Crawford}
  et~al.}{2010}]{2010SPIE.7737E..25C}
{Crawford} S.~M.,  et~al., 2010, in {Silva} D.~R.,  {Peck} A.~B.,   {Soifer}
  B.~T.,  eds,  Society of Photo-Optical Instrumentation Engineers (SPIE)
  Conference Series Vol. 7737, Observatory Operations: Strategies, Processes,
  and Systems III. p. 773725, \mn@doi{10.1117/12.857000}

\bibitem[\protect\citeauthoryear{{Cropper}}{{Cropper}}{1990}]{1990SSRv...54..195C}
{Cropper} M.,  1990, \mn@doi [\ssr] {10.1007/BF00177799}, \href
  {https://ui.adsabs.harvard.edu/abs/1990SSRv...54..195C} {54, 195}

\bibitem[\protect\citeauthoryear{{Evans} et~al.,}{{Evans}
  et~al.}{2009}]{2009MNRAS.397.1177E}
{Evans} P.~A.,  et~al., 2009, \mn@doi [\mnras]
  {10.1111/j.1365-2966.2009.14913.x}, \href
  {https://ui.adsabs.harvard.edu/abs/2009MNRAS.397.1177E} {397, 1177}

\bibitem[\protect\citeauthoryear{{Gaia Collaboration} et~al.,}{{Gaia
  Collaboration} et~al.}{2016}]{2016A&A...595A...1G}
{Gaia Collaboration} et~al., 2016, \mn@doi [\aap]
  {10.1051/0004-6361/201629272}, \href
  {https://ui.adsabs.harvard.edu/abs/2016A&A...595A...1G} {595, A1}

\bibitem[\protect\citeauthoryear{{Gaia Collaboration} et~al.,}{{Gaia
  Collaboration} et~al.}{2021}]{2021A&A...649A...1G}
{Gaia Collaboration} et~al., 2021, \mn@doi [\aap]
  {10.1051/0004-6361/202039657}, \href
  {https://ui.adsabs.harvard.edu/abs/2021A&A...649A...1G} {649, A1}

\bibitem[\protect\citeauthoryear{Garlick}{Garlick}{1993}]{Garlick_1993}
Garlick M.~A.,  1993, PhD thesis, University College London

\bibitem[\protect\citeauthoryear{{Ghosh} \& {Lamb}}{{Ghosh} \&
  {Lamb}}{1978}]{1978ApJ...223L..83G}
{Ghosh} P.,  {Lamb} F.~K.,  1978, \mn@doi [\apjl] {10.1086/182734}, \href
  {https://ui.adsabs.harvard.edu/abs/1978ApJ...223L..83G} {223, L83}

\bibitem[\protect\citeauthoryear{{HI4PI Collaboration} et~al.,}{{HI4PI
  Collaboration} et~al.}{2016}]{2016A&A...594A.116H}
{HI4PI Collaboration} et~al., 2016, \mn@doi [\aap]
  {10.1051/0004-6361/201629178}, \href
  {https://ui.adsabs.harvard.edu/abs/2016A&A...594A.116H} {594, A116}

\bibitem[\protect\citeauthoryear{{Halpern}}{{Halpern}}{2022}]{2022ApJ...934..123H}
{Halpern} J.~P.,  2022, \mn@doi [\apj] {10.3847/1538-4357/ac7d50}, \href
  {https://ui.adsabs.harvard.edu/abs/2022ApJ...934..123H} {934, 123}

\bibitem[\protect\citeauthoryear{{Halpern} \& {Thorstensen}}{{Halpern} \&
  {Thorstensen}}{2015}]{2015AJ....150..170H}
{Halpern} J.~P.,  {Thorstensen} J.~R.,  2015, \mn@doi [\aj]
  {10.1088/0004-6256/150/6/170}, \href
  {https://ui.adsabs.harvard.edu/abs/2015AJ....150..170H} {150, 170}

\bibitem[\protect\citeauthoryear{{Haswell}, {Patterson}, {Thorstensen},
  {Hellier}  \& {Skillman}}{{Haswell} et~al.}{1997}]{1997ApJ...476..847H}
{Haswell} C.~A.,  {Patterson} J.,  {Thorstensen} J.~R.,  {Hellier} C.,
  {Skillman} D.~R.,  1997, \mn@doi [\apj] {10.1086/303630}, \href
  {https://ui.adsabs.harvard.edu/abs/1997ApJ...476..847H} {476, 847}

\bibitem[\protect\citeauthoryear{{Hellier}}{{Hellier}}{2001}]{2001cvs..book.....H}
{Hellier} C.,  2001, {Cataclysmic Variable Stars}.
Springer, London

\bibitem[\protect\citeauthoryear{{Hellier}, {Mason}  \& {Mittaz}}{{Hellier}
  et~al.}{1991}]{1991MNRAS.248P...5H}
{Hellier} C.,  {Mason} K.~O.,   {Mittaz} J.~P.~D.,  1991, \mn@doi [\mnras]
  {10.1093/mnras/248.1.5P}, \href
  {https://ui.adsabs.harvard.edu/abs/1991MNRAS.248P...5H} {248, 5P}

\bibitem[\protect\citeauthoryear{{Hellier}, {Garlick}  \& {Mason}}{{Hellier}
  et~al.}{1993}]{1993MNRAS.260..299H}
{Hellier} C.,  {Garlick} M.~A.,   {Mason} K.~O.,  1993, \mn@doi [\mnras]
  {10.1093/mnras/260.2.299}, \href
  {https://ui.adsabs.harvard.edu/abs/1993MNRAS.260..299H} {260, 299}

\bibitem[\protect\citeauthoryear{{Hellier}, {Wynn}  \& {Buckley}}{{Hellier}
  et~al.}{2002}]{2002MNRAS.333...84H}
{Hellier} C.,  {Wynn} G.~A.,   {Buckley} D. A.~H.,  2002, \mn@doi [\mnras]
  {10.1046/j.1365-8711.2002.05381.x}, \href
  {https://ui.adsabs.harvard.edu/abs/2002MNRAS.333...84H} {333, 84}

\bibitem[\protect\citeauthoryear{{Joshi}, {Pandey}, {Singh}  \&
  {Agrawal}}{{Joshi} et~al.}{2016}]{2016ApJ...830...56J}
{Joshi} A.,  {Pandey} J.~C.,  {Singh} K.~P.,   {Agrawal} P.~C.,  2016, \mn@doi
  [\apj] {10.3847/0004-637X/830/2/56}, \href
  {https://ui.adsabs.harvard.edu/abs/2016ApJ...830...56J} {830, 56}

\bibitem[\protect\citeauthoryear{{Kemp}, {Patterson}, {Thorstensen}, {Fried},
  {Skillman}  \& {Billings}}{{Kemp} et~al.}{2002}]{2002PASP..114..623K}
{Kemp} J.,  {Patterson} J.,  {Thorstensen} J.~R.,  {Fried} R.~E.,  {Skillman}
  D.~R.,   {Billings} G.,  2002, \mn@doi [\pasp] {10.1086/341686}, \href
  {https://ui.adsabs.harvard.edu/abs/2002PASP..114..623K} {114, 623}

\bibitem[\protect\citeauthoryear{{King} \& {Lasota}}{{King} \&
  {Lasota}}{1991}]{1991ApJ...378..674K}
{King} A.~R.,  {Lasota} J.-P.,  1991, \mn@doi [\apj] {10.1086/170467}, \href
  {https://ui.adsabs.harvard.edu/abs/1991ApJ...378..674K} {378, 674}

\bibitem[\protect\citeauthoryear{{King} \& {Wynn}}{{King} \&
  {Wynn}}{1999}]{1999MNRAS.310..203K}
{King} A.~R.,  {Wynn} G.~A.,  1999, \mn@doi [\mnras]
  {10.1046/j.1365-8711.1999.02974.x}, \href
  {https://ui.adsabs.harvard.edu/abs/1999MNRAS.310..203K} {310, 203}

\bibitem[\protect\citeauthoryear{{King}, {Mouchet}  \& {Lasota}}{{King}
  et~al.}{1991}]{1991sepa.conf..213K}
{King} A.~R.,  {Mouchet} M.,   {Lasota} J.~P.,  1991, in {Bertout} C.,
  {Collin-Souffrin} S.,   {Lasota} J.~P.,  eds, IAU Colloq. 129: The 6th
  Institute d'Astrophysique de Paris (IAP) Meeting: Structure and Emission
  Properties of Accretion Disks. p.~213

\bibitem[\protect\citeauthoryear{{Knigge}, {Baraffe}  \& {Patterson}}{{Knigge}
  et~al.}{2011}]{2011ApJS..194...28K}
{Knigge} C.,  {Baraffe} I.,   {Patterson} J.,  2011, \mn@doi [\apjs]
  {10.1088/0067-0049/194/2/28}, \href
  {https://ui.adsabs.harvard.edu/abs/2011ApJS..194...28K} {194, 28}

\bibitem[\protect\citeauthoryear{{Kobulnicky}, {Nordsieck}, {Burgh}, {Smith},
  {Percival}, {Williams}  \& {O'Donoghue}}{{Kobulnicky}
  et~al.}{2003}]{2003SPIE.4841.1634K}
{Kobulnicky} H.~A.,  {Nordsieck} K.~H.,  {Burgh} E.~B.,  {Smith} M.~P.,
  {Percival} J.~W.,  {Williams} T.~B.,   {O'Donoghue} D.,  2003, in {Iye} M.,
  {Moorwood} A. F.~M.,  eds,  Society of Photo-Optical Instrumentation
  Engineers (SPIE) Conference Series Vol. 4841, Instrument Design and
  Performance for Optical/Infrared Ground-based Telescopes. pp 1634--1644,
  \mn@doi{10.1117/12.460315}

\bibitem[\protect\citeauthoryear{{Lamb} \& {Patterson}}{{Lamb} \&
  {Patterson}}{1983}]{1983ASSL..101..229L}
{Lamb} D.~Q.,  {Patterson} J.,  1983, in {Livio} M.,  {Shaviv} G.,  eds,
  Astrophysics and Space Science Library Vol. 101, IAU Colloq. 72: Cataclysmic
  Variables and Related Objects. pp 229--236,
  \mn@doi{10.1007/978-94-009-7118-9_29}

\bibitem[\protect\citeauthoryear{{Lang}, {Hogg}, {Mierle}, {Blanton}  \&
  {Roweis}}{{Lang} et~al.}{2010}]{2010AJ....139.1782L}
{Lang} D.,  {Hogg} D.~W.,  {Mierle} K.,  {Blanton} M.,   {Roweis} S.,  2010,
  \mn@doi [\aj] {10.1088/0004-6256/139/5/1782}, \href
  {https://ui.adsabs.harvard.edu/abs/2010AJ....139.1782L} {139, 1782}

\bibitem[\protect\citeauthoryear{{Littlefield} et~al.,}{{Littlefield}
  et~al.}{2023}]{2023ApJ...943L..24L}
{Littlefield} C.,  et~al., 2023, \mn@doi [\apjl] {10.3847/2041-8213/acaf04},
  \href {https://ui.adsabs.harvard.edu/abs/2023ApJ...943L..24L} {943, L24}

\bibitem[\protect\citeauthoryear{{Lopes de Oliveira}, {Bruch}, {Rodrigues},
  {Oliveira}  \& {Mukai}}{{Lopes de Oliveira}
  et~al.}{2020}]{2020ApJ...898L..40L}
{Lopes de Oliveira} R.,  {Bruch} A.,  {Rodrigues} C.~V.,  {Oliveira} A.~S.,
  {Mukai} K.,  2020, \mn@doi [\apjl] {10.3847/2041-8213/aba618}, \href
  {https://ui.adsabs.harvard.edu/abs/2020ApJ...898L..40L} {898, L40}

\bibitem[\protect\citeauthoryear{{Marchesini} et~al.,}{{Marchesini}
  et~al.}{2019}]{2019Ap&SS.364..153M}
{Marchesini} E.~J.,  et~al., 2019, \mn@doi [\apss] {10.1007/s10509-019-3642-9},
  \href {https://ui.adsabs.harvard.edu/abs/2019Ap&SS.364..153M} {364, 153}

\bibitem[\protect\citeauthoryear{{Mhlahlo}, {Buckley}, {Dhillon}, {Potter},
  {Warner}  \& {Woudt}}{{Mhlahlo} et~al.}{2007}]{2007MNRAS.378..211M}
{Mhlahlo} N.,  {Buckley} D.~A.~H.,  {Dhillon} V.~S.,  {Potter} S.~B.,  {Warner}
  B.,   {Woudt} P.~A.,  2007, \mn@doi [\mnras]
  {10.1111/j.1365-2966.2007.11762.x}, \href
  {https://ui.adsabs.harvard.edu/abs/2007MNRAS.378..211M} {378, 211}

\bibitem[\protect\citeauthoryear{{Mukai}}{{Mukai}}{1999}]{1999ASPC..157...33M}
{Mukai} K.,  1999, in {Hellier} C.,  {Mukai} K.,  eds,  Astronomical Society of
  the Pacific Conference Series Vol. 157, Annapolis Workshop on Magnetic
  Cataclysmic Variables. p.~33

\bibitem[\protect\citeauthoryear{{Myers} et~al.,}{{Myers}
  et~al.}{2017}]{2017PASP..129d4204M}
{Myers} G.,  et~al., 2017, \mn@doi [\pasp] {10.1088/1538-3873/aa54a8}, \href
  {https://ui.adsabs.harvard.edu/abs/2017PASP..129d4204M} {129, 044204}

\bibitem[\protect\citeauthoryear{{NASA High Energy Astrophysics Science Archive
  Research Center (Heasarc)}}{{NASA High Energy Astrophysics Science Archive
  Research Center (Heasarc)}}{2014}]{2014ascl.soft08004N}
{NASA High Energy Astrophysics Science Archive Research Center (Heasarc)} 2014,
  {HEAsoft: Unified Release of FTOOLS and XANADU}, Astrophysics Source Code
  Library, record ascl:1408.004 (\mn@eprint {ascl} {1408.004})

\bibitem[\protect\citeauthoryear{{Norton}, {McHardy}, {Lehto}  \&
  {Watson}}{{Norton} et~al.}{1992}]{1992MNRAS.258..697N}
{Norton} A.~J.,  {McHardy} I.~M.,  {Lehto} H.~J.,   {Watson} M.~G.,  1992,
  \mn@doi [\mnras] {10.1093/mnras/258.4.697}, \href
  {https://ui.adsabs.harvard.edu/abs/1992MNRAS.258..697N} {258, 697}

\bibitem[\protect\citeauthoryear{{Norton}, {Somerscales}, {Parker}, {Wynn}  \&
  {West}}{{Norton} et~al.}{2004a}]{2004RMxAC..20..138N}
{Norton} A.~J.,  {Somerscales} R.~V.,  {Parker} T.~L.,  {Wynn} G.~A.,   {West}
  R.~G.,  2004a, in {Tovmassian} G.,  {Sion} E.,  eds,  Revista Mexicana de
  Astronomia y Astrofisica Conference Series Vol. 20, Revista Mexicana de
  Astronomia y Astrofisica Conference Series. pp 138--139

\bibitem[\protect\citeauthoryear{{Norton}, {Wynn}  \& {Somerscales}}{{Norton}
  et~al.}{2004b}]{2004ApJ...614..349N}
{Norton} A.~J.,  {Wynn} G.~A.,   {Somerscales} R.~V.,  2004b, \mn@doi [\apj]
  {10.1086/423333}, \href
  {https://ui.adsabs.harvard.edu/abs/2004ApJ...614..349N} {614, 349}

\bibitem[\protect\citeauthoryear{{Norton}, {Butters}, {Parker}  \&
  {Wynn}}{{Norton} et~al.}{2008}]{2008ApJ...672..524N}
{Norton} A.~J.,  {Butters} O.~W.,  {Parker} T.~L.,   {Wynn} G.~A.,  2008,
  \mn@doi [\apj] {10.1086/523932}, \href
  {https://ui.adsabs.harvard.edu/abs/2008ApJ...672..524N} {672, 524}

\bibitem[\protect\citeauthoryear{{O'Donoghue} et~al.,}{{O'Donoghue}
  et~al.}{2006}]{2006MNRAS.372..151O}
{O'Donoghue} D.,  et~al., 2006, \mn@doi [\mnras]
  {10.1111/j.1365-2966.2006.10834.x}, \href
  {https://ui.adsabs.harvard.edu/abs/2006MNRAS.372..151O} {372, 151}

\bibitem[\protect\citeauthoryear{{Ohsuga}, {Mineshige}, {Mori}  \&
  {Umemura}}{{Ohsuga} et~al.}{2002}]{2002ApJ...574..315O}
{Ohsuga} K.,  {Mineshige} S.,  {Mori} M.,   {Umemura} M.,  2002, \mn@doi [\apj]
  {10.1086/340798}, \href
  {https://ui.adsabs.harvard.edu/abs/2002ApJ...574..315O} {574, 315}

\bibitem[\protect\citeauthoryear{{Patterson}}{{Patterson}}{1999}]{1999dicb.conf...61P}
{Patterson} J.,  1999, in {Mineshige} S.,  {Wheeler} J.~C.,  eds, Disk
  Instabilities in Close Binary Systems. p.~61

\bibitem[\protect\citeauthoryear{{Patterson} et~al.,}{{Patterson}
  et~al.}{2004}]{2004PASP..116..516P}
{Patterson} J.,  et~al., 2004, \mn@doi [\pasp] {10.1086/421034}, \href
  {https://ui.adsabs.harvard.edu/abs/2004PASP..116..516P} {116, 516}

\bibitem[\protect\citeauthoryear{{Postma} \& {Leahy}}{{Postma} \&
  {Leahy}}{2017}]{2017PASP..129k5002P}
{Postma} J.~E.,  {Leahy} D.,  2017, \mn@doi [\pasp] {10.1088/1538-3873/aa8800},
  \href {https://ui.adsabs.harvard.edu/abs/2017PASP..129k5002P} {129, 115002}

\bibitem[\protect\citeauthoryear{{Pretorius} \& {Mukai}}{{Pretorius} \&
  {Mukai}}{2014}]{2014MNRAS.442.2580P}
{Pretorius} M.~L.,  {Mukai} K.,  2014, \mn@doi [\mnras] {10.1093/mnras/stu990},
  \href {https://ui.adsabs.harvard.edu/abs/2014MNRAS.442.2580P} {442, 2580}

\bibitem[\protect\citeauthoryear{Rawat, Pandey, Joshi, Scaringi  \&
  Yadava}{Rawat et~al.}{2022}]{Rawat:2022onw}
Rawat N.,  Pandey J.~C.,  Joshi A.,  Scaringi S.,   Yadava U.,  2022, \mn@doi
  [Mon. Not. Roy. Astron. Soc.] {10.1093/mnras/stac2723}, 517, 1667

\bibitem[\protect\citeauthoryear{{Ricker} et~al.,}{{Ricker}
  et~al.}{2015}]{2015JATIS...1a4003R}
{Ricker} G.~R.,  et~al., 2015, \mn@doi [Journal of Astronomical Telescopes,
  Instruments, and Systems] {10.1117/1.JATIS.1.1.014003}, \href
  {https://ui.adsabs.harvard.edu/abs/2015JATIS...1a4003R} {1, 014003}

\bibitem[\protect\citeauthoryear{{Rosen}}{{Rosen}}{1987}]{1987PhDT.......130R}
{Rosen} S.~R.,  1987, PhD thesis, University College London, Mullard Space
  Science Laboratory

\bibitem[\protect\citeauthoryear{{Rosen}, {Mason}  \& {Cordova}}{{Rosen}
  et~al.}{1988}]{1988MNRAS.231..549R}
{Rosen} S.~R.,  {Mason} K.~O.,   {Cordova} F.~A.,  1988, \mn@doi [\mnras]
  {10.1093/mnras/231.3.549}, \href
  {https://ui.adsabs.harvard.edu/abs/1988MNRAS.231..549R} {231, 549}

\bibitem[\protect\citeauthoryear{{Scaringi} et~al.,}{{Scaringi}
  et~al.}{2010}]{2010MNRAS.401.2207S}
{Scaringi} S.,  et~al., 2010, \mn@doi [\mnras]
  {10.1111/j.1365-2966.2009.15826.x}, \href
  {https://ui.adsabs.harvard.edu/abs/2010MNRAS.401.2207S} {401, 2207}

\bibitem[\protect\citeauthoryear{{Schwarz}, {Schwope}, {Staude}, {Rau},
  {Hasinger}, {Urrutia}  \& {Motch}}{{Schwarz}
  et~al.}{2007}]{2007A&A...473..511S}
{Schwarz} R.,  {Schwope} A.~D.,  {Staude} A.,  {Rau} A.,  {Hasinger} G.,
  {Urrutia} T.,   {Motch} C.,  2007, \mn@doi [\aap]
  {10.1051/0004-6361:20077684}, \href
  {https://ui.adsabs.harvard.edu/abs/2007A&A...473..511S} {473, 511}

\bibitem[\protect\citeauthoryear{{Schwope}, {Mantel}  \& {Horne}}{{Schwope}
  et~al.}{1997}]{1997A&A...319..894S}
{Schwope} A.~D.,  {Mantel} K.~H.,   {Horne} K.,  1997, \aap, \href
  {https://ui.adsabs.harvard.edu/abs/1997A&A...319..894S} {319, 894}

\bibitem[\protect\citeauthoryear{{Silber}}{{Silber}}{1992}]{1992PhDT.......119S}
{Silber} A.~D.,  1992, PhD thesis, Massachusetts Institute of Technology

\bibitem[\protect\citeauthoryear{{Singh} et~al.,}{{Singh}
  et~al.}{2017}]{2017JApA...38...29S}
{Singh} K.~P.,  et~al., 2017, \mn@doi [Journal of Astrophysics and Astronomy]
  {10.1007/s12036-017-9448-7}, \href
  {https://ui.adsabs.harvard.edu/abs/2017JApA...38...29S} {38, 29}

\bibitem[\protect\citeauthoryear{{Singh} et~al.,}{{Singh}
  et~al.}{2021}]{2021JApA...42...83S}
{Singh} K.~P.,  et~al., 2021, \mn@doi [Journal of Astrophysics and Astronomy]
  {10.1007/s12036-021-09756-w}, \href
  {https://ui.adsabs.harvard.edu/abs/2021JApA...42...83S} {42, 83}

\bibitem[\protect\citeauthoryear{{Smith} \& {Dhillon}}{{Smith} \&
  {Dhillon}}{1998}]{1998MNRAS.301..767S}
{Smith} D.~A.,  {Dhillon} V.~S.,  1998, \mn@doi [\mnras]
  {10.1046/j.1365-8711.1998.02065.x}, \href
  {https://ui.adsabs.harvard.edu/abs/1998MNRAS.301..767S} {301, 767}

\bibitem[\protect\citeauthoryear{{Smithsonian Astrophysical
  Observatory}}{{Smithsonian Astrophysical
  Observatory}}{2000}]{2000ascl.soft03002S}
{Smithsonian Astrophysical Observatory} 2000, {SAOImage DS9: A utility for
  displaying astronomical images in the X11 window environment}, Astrophysics
  Source Code Library, record ascl:0003.002 (\mn@eprint {ascl} {0003.002})

\bibitem[\protect\citeauthoryear{{Southworth}, {G{\"a}nsicke}, {Marsh}, {de
  Martino}  \& {Aungwerojwit}}{{Southworth} et~al.}{2007}]{2007MNRAS.378..635S}
{Southworth} J.,  {G{\"a}nsicke} B.~T.,  {Marsh} T.~R.,  {de Martino} D.,
  {Aungwerojwit} A.,  2007, \mn@doi [\mnras]
  {10.1111/j.1365-2966.2007.11796.x}, \href
  {https://ui.adsabs.harvard.edu/abs/2007MNRAS.378..635S} {378, 635}

\bibitem[\protect\citeauthoryear{{Tandon} et~al.,}{{Tandon}
  et~al.}{2020}]{2020AJ....159..158T}
{Tandon} S.~N.,  et~al., 2020, \mn@doi [\aj] {10.3847/1538-3881/ab72a3}, \href
  {https://ui.adsabs.harvard.edu/abs/2020AJ....159..158T} {159, 158}

\bibitem[\protect\citeauthoryear{{Thorstensen} \& {Halpern}}{{Thorstensen} \&
  {Halpern}}{2013}]{2013AJ....146..107T}
{Thorstensen} J.~R.,  {Halpern} J.,  2013, \mn@doi [\aj]
  {10.1088/0004-6256/146/5/107}, \href
  {https://ui.adsabs.harvard.edu/abs/2013AJ....146..107T} {146, 107}

\bibitem[\protect\citeauthoryear{{Tody}}{{Tody}}{1986}]{1986SPIE..627..733T}
{Tody} D.,  1986, in {Crawford} D.~L.,  ed.,  Society of Photo-Optical
  Instrumentation Engineers (SPIE) Conference Series Vol. 627, Instrumentation
  in astronomy VI. p.~733, \mn@doi{10.1117/12.968154}

\bibitem[\protect\citeauthoryear{{Tonry} et~al.,}{{Tonry}
  et~al.}{2018}]{2018PASP..130f4505T}
{Tonry} J.~L.,  et~al., 2018, \mn@doi [\pasp]
  {10.1088/1538-3873/aabadf10.48550/arXiv.1802.00879}, \href
  {https://ui.adsabs.harvard.edu/abs/2018PASP..130f4505T} {130, 064505}

\bibitem[\protect\citeauthoryear{{Warner}}{{Warner}}{1995}]{1995cvs..book.....W}
{Warner} B.,  1995, {Cataclysmic variable stars}.
Cataclysmic Variable Stars

\bibitem[\protect\citeauthoryear{{Williams}}{{Williams}}{1980}]{1980ApJ...235..939W}
{Williams} R.~E.,  1980, \mn@doi [\apj] {10.1086/157698}, \href
  {https://ui.adsabs.harvard.edu/abs/1980ApJ...235..939W} {235, 939}

\bibitem[\protect\citeauthoryear{{Woudt} et~al.,}{{Woudt}
  et~al.}{2012}]{2012MNRAS.427.1004W}
{Woudt} P.~A.,  et~al., 2012, \mn@doi [\mnras]
  {10.1111/j.1365-2966.2012.22010.x}, \href
  {https://ui.adsabs.harvard.edu/abs/2012MNRAS.427.1004W} {427, 1004}

\bibitem[\protect\citeauthoryear{{Wynn}}{{Wynn}}{2000}]{2000NewAR..44...75W}
{Wynn} G.~A.,  2000, \mn@doi [\nar] {10.1016/S1387-6473(00)00017-8}, \href
  {https://ui.adsabs.harvard.edu/abs/2000NewAR..44...75W} {44, 75}

\bibitem[\protect\citeauthoryear{{Wynn}}{{Wynn}}{2001}]{2001LNP...573..155W}
{Wynn} G.,  2001, in {Boffin} H.~M.~J.,  {Steeghs} D.,   {Cuypers} J.,  eds, ,
  Vol.~573, Astrotomography, Indirect Imaging Methods in Observational
  Astronomy.
Springer, Berlin, Heidelberg, p.~155

\bibitem[\protect\citeauthoryear{{Wynn} \& {King}}{{Wynn} \&
  {King}}{1992}]{1992MNRAS.255...83W}
{Wynn} G.~A.,  {King} A.~R.,  1992, \mn@doi [\mnras] {10.1093/mnras/255.1.83},
  \href {https://ui.adsabs.harvard.edu/abs/1992MNRAS.255...83W} {255, 83}

\bibitem[\protect\citeauthoryear{{Wynn}, {King}  \& {Horne}}{{Wynn}
  et~al.}{1997}]{1997MNRAS.286..436W}
{Wynn} G.~A.,  {King} A.~R.,   {Horne} K.,  1997, \mn@doi [\mnras]
  {10.1093/mnras/286.2.436}, \href
  {https://ui.adsabs.harvard.edu/abs/1997MNRAS.286..436W} {286, 436}

\bibitem[\protect\citeauthoryear{{Zhang}, {Robinson}, {Stiening}  \&
  {Horne}}{{Zhang} et~al.}{1995}]{1995ApJ...454..447Z}
{Zhang} E.,  {Robinson} E.~L.,  {Stiening} R.~F.,   {Horne} K.,  1995, \mn@doi
  [\apj] {10.1086/176496}, \href
  {https://ui.adsabs.harvard.edu/abs/1995ApJ...454..447Z} {454, 447}

\bibitem[\protect\citeauthoryear{{Zhilkin}, {Bisikalo}  \& {Mason}}{{Zhilkin}
  et~al.}{2012}]{2012ARep...56..257Z}
{Zhilkin} A.~G.,  {Bisikalo} D.~V.,   {Mason} P.~A.,  2012, \mn@doi [Astronomy
  Reports] {10.1134/S1063772912040087}, \href
  {https://ui.adsabs.harvard.edu/abs/2012ARep...56..257Z} {56, 257}

\bibitem[\protect\citeauthoryear{{de Jager}, {Meintjes}, {O'Donoghue}  \&
  {Robinson}}{{de Jager} et~al.}{1994}]{1994MNRAS.267..577D}
{de Jager} O.~C.,  {Meintjes} P.~J.,  {O'Donoghue} D.,   {Robinson} E.~L.,
  1994, \mn@doi [\mnras] {10.1093/mnras/267.3.577}, \href
  {https://ui.adsabs.harvard.edu/abs/1994MNRAS.267..577D} {267, 577}

\makeatother
\end{thebibliography}










\bsp	
\label{lastpage}
\end{document}